\documentclass[
reprint,
amsmath,assume,
aps
]{revtex4-2}

\usepackage[dvipsnames]{xcolor}
\usepackage{graphicx} 
\usepackage{subcaption}
\usepackage{derivative}
\usepackage{dcolumn} 
\usepackage{booktabs}
\usepackage{wrapfig}
\usepackage{placeins}
\usepackage{lipsum} 
\usepackage{float}
\usepackage{bm} 
\usepackage{hyperref} 

\usepackage{xcolor}
\usepackage{xparse}
\usepackage{amsfonts}
\usepackage{physics}
\usepackage{parskip} 
\usepackage{etoolbox} 

\AtEndEnvironment{figure}{\ignorespacesafterend}
\AtEndEnvironment{equation}{\ignorespacesafterend}

\usepackage{graphicx}
\usepackage{bm}

\begin{document}

\preprint{APS/123-QED}
\title{Anderson localization via Peierls phase modulation}

\author{Arpita Goswami}
\thanks{These authors contributed equally}
\author{Pallabi Chatterjee}
\thanks{These authors contributed equally}
\author{Ranjan Modak}
\thanks{Contact author: ranjan@iittp.ac.in}
\author{Shaon Sahoo}
\thanks{Contact author: shaon@iittp.ac.in}
\affiliation{Department of Physics, Indian Institute of Technology Tirupati, India, 517619}

\begin{abstract}
We investigate a two-leg ladder system subjected to an external magnetic field. In the absence of a magnetic field, the system is described by a clean tight-binding model, with no disorder in either the on-site potential or the hopping amplitudes. The effect of magnetic field in this system is studied by introducing the Peierls phases in the hopping amplitudes along a leg (appropriate when the Landau gauge is chosen). 
For a uniform magnetic field, characterized by a constant Peierls phase, we find that all eigenstates remain delocalized. In contrast, random Peierls phases, representing a random magnetic field, lead to complete localization of the eigenstates. We further show that a quasiperiodic modulation of the Peierls phase can drive a transition from a fully delocalized to a fully localized phase upon tuning the quasiperiodicity. For a two-parameter quasiperiodic Peierls phase, varying analogously to a generalized Aubry–Andr\'{e}–type potential, we construct the phase diagram of the system. The phase diagram exhibits regions of delocalized and localized phases, separated by intermediate regimes of mixed phase. We also perform a semiclassical analysis that qualitatively yields a similar phase diagram, capturing the localization transition. Our results demonstrate a mechanism for controlling transport properties via the Peierls phase engineering.
\end{abstract}

\maketitle

\section{Introduction}\label{sec:introduction}

In one- and two-dimensional non-interacting systems, arbitrarily small disorder can suppress electronic transport due to Anderson localization (AL)~\cite{AL_1, AL_2}. In such cases, all single-particle states become exponentially localized even in the presence of infinitesimal disorder~\cite{SL_1, SL_2}. However, in three-dimensional systems it has been established that, for finite disorder strength, extended and localized states can coexist, separated by a critical energy $E_c$, known as the mobility edge (ME)~\cite{pr_sc_2}. 

Introducing interactions into a system exhibiting Anderson localization can give rise to a many-body localized (MBL) phase that fails to thermalize~\cite{MBL1,MBL2,MBL3,MBL4,MBL_rev1,nandkishore2015many,sierant25}, thereby violating the Eigenstate Thermalization Hypothesis (ETH), which generally describes thermalization in isolated quantum systems~\cite{ETH1, ETH2, rigol2012alternatives}. However, recent studies have highlighted the avalanche instability of the MBL phase due to rare thermal regions in the system. These regions can act as local thermal baths for nearby localized regions, leading to the growth of entanglement and potentially destabilizing the MBL phase~\cite{MBL_breakdown1, MBL_breakdown2, vidmar.2020,sierant25}.

Unlike in non-interacting disordered system, where the metal--insulator (MI) transition occurs only in dimension $d\geq 3$, systems with quasiperiodic disorder can exhibit a similar transition even in low dimensions. A paradigmatic example is the self-dual Aubry--Andr\'e (AA) model, which shows a sharp MI transition at a critical quasiperiodic potential strength, beyond (before) which all eigenstates become localized (extended)~\cite{QP_1,QP_2,QP_3,QP_4,QP_5,QP_6,QP_7,QP_8,QP_9,QP_10,QP_11}. Extensions of the AA model further reveal the presence of mobility edges, leading to a finite parameter regime where extended and localized states coexist~\cite{ME_1,ME_2,ME_3,ME_4,ME_5,ME_6}. These predictions have been experimentally explored in several platforms, including ultracold atoms~\cite{EXP_1,EXP_2,EXP_3,EXP_4,EXP_5,EXP_6,EXP_7,EXP_8,EXP_9}, photonic crystals~\cite{expt_1,expt_2,expt_3,expt_4}, electrical circuits~\cite{electric_circuit}, optical cavities, and superconducting circuits~\cite{sc_qubit_1,sc_qubit_2}. Quasiperiodic systems can also host a critical phase characterized by states that are extended yet nonergodic and exhibit scale-invariant properties, leading to unusual spectral statistics, fractal structures, and distinct dynamical behavior~\cite{critical_1,critical_2,critical_3,critical_4,nonergodic_1,spec_stat_1,spec_stat_3,spec_stat_4,spec_stat_5,frac_1,frac_2,frac_3,dyn_1,dyn_2}. In the presence of interactions, such phases may give rise to a many-body critical regime that lies between the thermal and many-body localized phases~\cite{nonergodic_1,nem_modak,nem_soumi,nem_sriram,deng2017many,MBL2}.
Another intriguing phenomenon in quasiperiodic systems is the reentrant localization transition, where increasing disorder initially localizes the system, followed by a partial recovery of extended states at intermediate disorder strengths, before eventual relocalization at stronger disorder~\cite{pr_sc_1, QP_4}. In one-dimensional systems, this behavior typically arises from the competition between dimerization and quasiperiodic disorder~\cite{QP_4, dyn_2, Karmakar2025}, while in quasi-one-dimensional ladder systems it can emerge from the interplay between interchain hopping and onsite potential modulation~\cite{goswami25}. 

In parallel with disorder and quasiperiodic on-site potentials, magnetic fields have long served as a central tool for uncovering fundamental quantum phenomena in condensed matter systems. One of the most striking examples is the quantum Hall effect~\cite{Hall_1,Hall_2,Hall_3,Hall_4,Hall_5}, where a strong perpendicular magnetic field quantizes electronic motion into Landau levels, leading to precisely quantized Hall conductance and topologically protected edge states that have been extensively studied both theoretically and experimentally. The quantum Hall effect established that magnetic fields can fundamentally reorganize electronic spectra, generate topological invariants, and produce robust boundary modes that are insensitive to microscopic disorder. In two‑dimensional lattices, random flux has also been shown to drive topological transitions~\cite{2D_rndm1, 2D_rndm2, 2D_rndm3, 2D_rndm4}, highlighting the role of flux disorder in quantum systems. Closely related are magnetoresistance phenomena~\cite{mag_res_1,mag_res_2}, where the application of a magnetic field significantly modifies the transport properties of electronic systems.

In lattice systems, the effect of a magnetic field can be incorporated through the Peierls substitution~\cite{peierls}, which introduces complex phase factors in the hopping amplitudes and modifies quantum interference without adding explicit onsite disorder. Such phase engineering can significantly influence spectral and transport properties and may alter localization behavior~\cite{zigzag_flux}. A paradigmatic example is the Hofstadter model~\cite{H_butterfly}, describing electrons on a two-dimensional lattice under a strong magnetic field, where the interplay between lattice periodicity and magnetic length leads to a fractal energy spectrum known as the Hofstadter butterfly. Experimental realizations of this effect typically rely on superlattice structures, such as graphene and semiconductor heterostructures, where enlarged lattice constants make the required magnetic flux experimentally accessible~\cite{HB_1,HB_2,HB_3}. The action of infinitesimal flux disorder on the localization properties of flat band states is studied in \cite{flat_band_disorder}.
Recently, reduced-dimensional implementations of the Hofstadter model, particularly in strip geometries, have attracted considerable attention~\cite{HB_4,HB_5,HB_6,HB_7}. Among them, two-leg ladder systems~\cite{chiral_ladder} provide a minimal yet nontrivial platform that retains plaquettes enclosing magnetic flux while restricting transverse motion. 

In this work, we aim to identify a minimal low-dimensional short-range model in which transport can be controlled by tuning an external magnetic field (or equivalently, the magnetic flux). In particular, we ask whether a spatially dependent magnetic field alone can drive a delocalization--localization (metal--insulator) transition in the single-particle spectrum without modifying the onsite potential. We answer this question affirmatively. Our analysis shows that such a transition cannot occur in a purely one-dimensional system; instead, at least a quasi-one-dimensional geometry, such as a ladder lattice, is required. Within this framework, we identify parameter regimes where the system is fully delocalized or fully localized, or it is in some intermediate or mixed phase.

The manuscript is organized as follows. In Sec.~\ref{sec2}, we discuss the theoretical setup, including the model and its phase diagram. The diagnostic tools are introduced in Sec.~\ref{diagn_tools}. We then present the numerical results in Sec.~\ref{sec: numerical analysis}, followed by the semiclassical analysis in Sec.~\ref{sec: semiclassical analysis}. Finally, we conclude in Sec.~\ref{conclusion}.

\begin{figure}[t]
\includegraphics[width=1\linewidth]{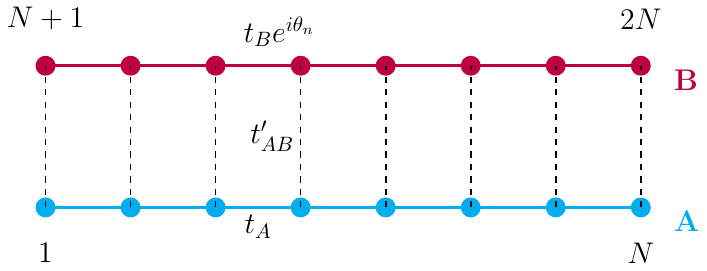}
    \caption{Schematic diagram of the model system, represented by Hamiltonian of Eq. \ref{ham_eqn}.}   \label{fig0}
\end{figure}
\section{Theoretical setup} \label{sec2}
\subsection{Model}
The main goal of this work is to investigate a minimal, low-dimensional experimentally realizable system
of non-interacting short-range fermions, in which the transport properties can be controlled through an external magnetic field. First, a natural choice is to consider a one-dimensional (1D) tight-binding model in the presence of an inhomogeneous magnetic field. Such a model can be described by the Hamiltonian

\begin{equation}
H_{1D}= -t \sum_{n} \left( e^{i\theta_n} c_{n+1}^{\dagger} c_n + e^{-i\theta_n} c_n^{\dagger} c_{n+1} \right),
\label{Eq: Hamil 1D}
\end{equation}
where $c_n$ ($c^{\dag}_n$) is fermionic annihilation (creation) operator, $t$ is the hopping amplitude and $\theta_n$ is the Peierls phase given by

\begin{equation}
\theta_n = \frac{e}{\hbar} \int_{x_n}^{x_{n+1}} \mathbf{A}\cdot d\mathbf{l},
\label{Eq: vector pot}
\end{equation}
where $x_n$ and $x_{n+1}$ denote the positions of the $n$th and $(n+1)$th lattice sites, respectively. Here $\mathbf{A}$ is the position-dependent vector potential, which can be tuned to generate different types of inhomogeneous magnetic fields.
 $H_{1D}$ might appear non-trivial at first glance, and one might expect that by varying $\theta_n$ it could be possible to induce localization or control transport in the system. However, it turns out that under open boundary conditions,  since the chain has no closed loops, the phases $\theta_n$
can be eliminated by performing a local gauge transformation of the fermionic operators,
\begin{equation}
c_n = e^{i\tilde{\theta}_n} d_n .\nonumber 
\end{equation}
Substituting this into the Hamiltonian $H_{1D}$ and choosing the gauge such that $\tilde{\theta}_{n+1} - \tilde{\theta}_n = \theta_n$,
the Hamiltonian (Eq.~\eqref{Eq: Hamil 1D}) reduces to

\begin{equation}
H_{1D} = -t \sum_n \left( d_{n+1}^{\dagger} d_n + d_n^{\dagger} d_{n+1} \right).\nonumber 
\end{equation}
Here $d_n$ and $d_n^{\dagger}$ are also fermionic annihilation and creation operators that obey the same anticommutation relations as $c_n$ and $c_n^{\dagger}$. Therefore, the transformed Hamiltonian is exactly identical to the standard nearest-neighbour tight-binding model. The eigenstates of this model are completely delocalized, and hence no localization–delocalization transition can occur in this system.

Hence, we focus on a quasi-one-dimensional model, namely a ladder system with two legs and rungs (see Fig.~\ref{fig0}). In contrast to a purely one-dimensional chain, a ladder geometry contains closed loops (plaquettes) formed by the two legs and the rungs. When Peierls phases are introduced in such a system, the sum of the phases around each plaquette corresponds to the magnetic flux threading that plaquette. This total phase cannot be eliminated by a gauge transformation, as was possible in the strictly one-dimensional case, and therefore becomes a physically observable quantity. As a result, the magnetic field modifies the band structure and transport properties of the ladder system.
To make our model more concrete, we consider the ladder lying in the $xy$ plane, with a magnetic field applied in the $z$ direction. We choose the Landau gauge 
$\mathbf{A} = (-By, 0, 0)$,
which satisfies $\mathbf{B} = \nabla \times \mathbf{A}$. B is the magnitude of the magnetic field (which is not
constant for the disorder/quasiperiodic cases).

The Peierls phase $\theta_n$ is computed using Eq.~\eqref{Eq: vector pot} as the line integral of the vector potential along a closed plaquette, yielding the magnetic flux threading each unit cell.
Using gauge freedom, we place the lower leg along the $y=0$ axis. As a result, the hopping along the lower leg (referred to as the A leg) does not acquire any phase, whereas particles accumulate a Peierls phase only when moving along the upper leg (the B leg). Consequently, the effect of the magnetic flux is incorporated through a spatially modulated Peierls phase attached to the hopping amplitudes in the upper arm of the ladder~\cite{sajid2020localization}.

The full Hamiltonian of the ladder system appears in the following equation:
\begin{equation}\label{ham_eqn}
\begin{split}
H &= H_A+H_B+H_{AB},~ \text{where}\\
H_A &= -t_A\sum_{m=1}^{N-1}(c_m^{\dagger}c_{m+1}+hc), \\
H_B &= -t_B\sum_{n=N+1}^{2N-1}(e^{i\theta_n}c_n^{\dagger}c_{n+1}+hc),~ \text{and}\\
H_{AB} &= -t'_{AB}\sum_{j=1}^N(c_j^{\dagger}c_{N+j}+hc).
\end{split}
\end{equation} We take $t_A=t_B=t'_{AB}=1$ and $\beta=\frac{\sqrt{5}-1}{2}$, 
and 
\begin{equation} \label{theta_n}
\theta_n=\frac{V \pi \cos{(2\pi\beta n + \phi)}}{1-\lambda \cos{(2\pi\beta n + \phi)}}+\theta. 
\end{equation}
Here $V$ and $\lambda$ are the two parameters associated with the Peierls phase $\theta_n$, and  $\theta$ and $\phi$ are the overall phase-shift factors. For site-ordering, we first take all A sites sequentially, then the B sites, continuing the same sequential ordering. We denote the length of A leg as $N$; consequently, the total size of the system is $2N$.  In our calculations, we use an open boundary condition.\\

We consider three distinct scenarios. 

(A) \textbf{Uniform magnetic flux:} We set $V=0$ and assume a constant Peierls phase, $\theta_n = \theta$, for all lattice sites $n$. In this case, the magnetic flux is spatially uniform throughout the system.

(B) \textbf{Random magnetic flux:} We allow the Peierls phase $\theta_n$ to be independently drawn from a uniform distribution in the interval $[0,2\pi)$ for each site $n$. This corresponds to a spatially disordered magnetic flux configuration.

(C) \textbf{Quasiperiodic magnetic flux:} We focus on the regime $V \neq 0$. In particular, we primarily restrict our analysis to the parameter range $0 < V \le 6$ and $0 \le \lambda < 1$. When necessary, physical observables are averaged over different realizations of $\phi$ and $\theta$ to ensure reliable statistical convergence.

\subsection{Phase diagram}
\begin{figure}[t]
    \centering
   \includegraphics[width=0.44\textwidth]{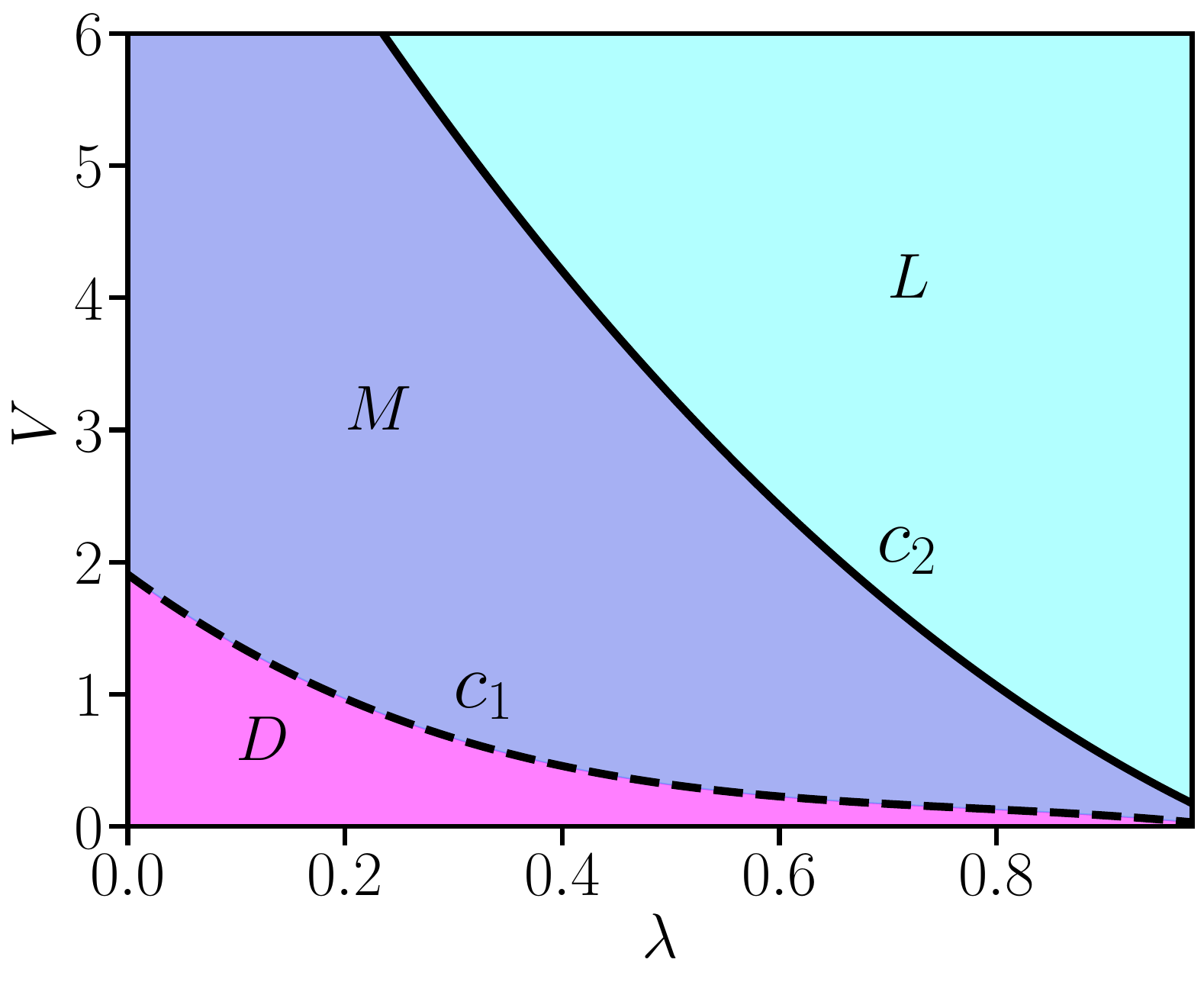}
    \caption{Schematic phase diagram corresponding to the quasiperiodic magnetic flux represented by Eq. \ref{theta_n}: $D$ = delocalized phase, $M$ = mixed (intermediate) phase, $L$ = localized phase. The $c_1$ and $c_2$ lines represent two crossover lines in the phase diagram.}
\label{phase_diag}
\end{figure}

The summary of our main results is presented in the schematic diagram of Fig. \ref{phase_diag}. It is constructed based on our results from the static and dynamical study of the system represented by Eqs. \ref{ham_eqn} and \ref{theta_n}. For uniform magnetic flux ($V=0$ in Eq. \ref{theta_n}), we have only the delocalized ($D$) phase. On the other hand, for random magnetic flux ($\theta_n$ chosen randomly), the system only exhibits the localized ($L$) phase. The study of $\langle NPR \rangle$ and $\langle IPR \rangle$ (see Sec. \ref{statics_tools} for definitions) for carefully chosen points (lines) in the $\lambda - V$ space reveals that the system exhibits three phases - delocalized ($D$), intermediate or mixed ($M$) and localized ($L$) phases. As shown in the phase diagram, two crossover lines, $c_1$ and $c_2$, separate $D$ and $M$ phases, and $M$ and $L$ phases, respectively. 

The study of single-particle dynamics reveals the values of the dynamical exponent ($\gamma$) and the saturation value of the mean squared displacement ($\sigma^2_{sat}$) in different phases (see Sec. \ref{dyna_tools} for definitions). It is found that the dynamics is ballistic ($\gamma\simeq 2$) in $D$ phase, super- or sub-diffusive  ($0<\gamma<2$) in $M$ phase and localized ($\gamma \simeq 0$) in $L$ phase. Interestingly, 
we also get the crossover line $c_2$ by studying the long-time saturation value of the variance ($\sigma^2_{sat}$) and its system size dependence.     

\section{Diagnostic tools} \label{diagn_tools}
We employ different diagnostic tools to analyze the localization properties of our model. In this section, we briefly discuss them. In the following discussion, $\ket{\xi_k}$ represents the $k$th eigenket of the system and $\ket{j}$ represents a standard site-basis ket which corresponds to a single particle at the $j$th site (and no particles at any other site). 

\subsection{Static tools} \label{statics_tools}
The inverse participation ratio (IPR) is known to measure the extent to which a state is localized. For the $k$th normalized eigenket, it is defined as \cite{ipr_1, ipr_2, ipr_3}:
\begin{equation} \label{ipr}
    IPR^{(k)} = \sum_{j=1}^{2N} |\braket{j}{\xi_k}|^4.
\end{equation}
We recall that $0\le IPR^{(k)}\le 1$. A higher IPR value, $IPR^{(k)} = O(1)$, indicates that the corresponding eigenket is localized, while a lower value ($\sim 1/N$) suggests that the corresponding state is delocalized.  

Sometimes, average IPR is calculated to understand overall localization properties of the system. It is defined as
\begin{equation} \label{ipr_av}
    \langle IPR \rangle = \frac{1}{2N}\sum_{k=1}^{2N} IPR^{(k)}. 
\end{equation}
A delocalized phase is characterized by $\langle IPR \rangle=0$ in the limit $N\to \infty$.

The participation ratio (PR) is a complementary, sometimes more convenient, measure of localization. We define the average PR in the following way:
\begin{equation}\label{pr_av}
    \langle PR \rangle = \frac{1}{2N} \sum_{k=1}^{2N} \frac{1}{IPR^{(k)}}.
\end{equation}
It may be noted here that, for the case of quasiperiodic flux, the quantity $\langle PR \rangle$ (or equivalently, $\langle IPR \rangle$) is averaged over 20 realization of the global phase $\phi$ or $\theta$ chosen uniformly from the interval $[0, 2\pi)$ (appears in the expression of $\theta_n$ in Eq. \ref{theta_n}) for better results. For the case of random magnetic flux, as discussed in Sec. \ref{ran_mag}, we average $\langle PR \rangle$ over 20 independent random realizations of $\theta_n$ drawn uniformly from $[0,2\pi)$. The corresponding averaged quantity is denoted by $\langle \overline{PR} \rangle$.

For a thermodynamically large system (i.e., $2N$ being large), the average PR is expected to follow the scaling behaviour \cite{pr_sc_1, pr_sc_2}: 
\begin{equation}\label{fsc_alph}
    \langle PR \rangle \sim N^\alpha,
\end{equation} 
where the exponent $\alpha$ represents the fractal nature of the system. This is a special case of a more general scaling relation, which can be used for studying the multifractal nature of the system: 
\begin{equation} \label{fractal_dim}
\frac{1}{2N}\sum_{k=1}^{2N}
\left[
\left(
\sum_{j=1}^{2N}
|\langle j \,|\, \xi_k \rangle|^{2q}
\right)^{-1}
\right]
\sim N^{D_q(q-1)}.
\end{equation}
Here, $D_q$ is the fractal dimension of the system in a given phase. Our special case, as presented in Eq. \ref{fsc_alph}, corresponds to $q=2$ and $D_q=\alpha$. The system is considered to be in the delocalized (localized) phase when $\alpha=1$ (0). The system is said to be in a fractal phase if $0<\alpha<1$. We point out here that the introduction of the fractal dimension $\alpha$ (or $D_q$) for a subsystem was done in a similar way in \cite{goswami25}. The current definition of the fractal dimension also closely parallels its definition via the scaling relation of the average inverse participation ratio (IPR) \cite{dyn_2,pr_sc_2}.

To gain more insight into the phases of a system, sometimes normalized participation ratio (NPR) is also calculated. For the $k$th eigenket, it is defined as 
\begin{equation}\label{npr}
    NPR^{(k)} = \frac{PR^{(k)}}{2N},
\end{equation}
where $PR^{(k)}=\frac{1}{IPR^{(k)}}$ (see Eq. \ref{ipr}). The average of this quantity over all eigenkets provides a good understanding of the overall localization behavior of a system. The average NPR is calculated following,
\begin{equation}\label{npr_av}
    \langle NPR\rangle = \frac{1}{2N}\sum_{k=1}^{2N} NPR^{(k)} = \frac{\langle PR \rangle}{2N},
\end{equation}
where $\langle PR \rangle$ is defined in Eq. \ref{pr_av}.
It is clear from Eq. \ref{fsc_alph} that, in the limit $N\to \infty$, $\langle NPR\rangle > 0$ for the delocalized phase and  $\langle NPR\rangle = 0$ for the localized and fractal phases. 

Following Refs. \cite{Li2020,QP_4} and other related works, we call a phase delocalized if $\langle IPR \rangle = 0$ and $\langle NPR \rangle > 0$, and designate a phase localized if $\langle IPR \rangle > 0$ and $\langle NPR \rangle = 0$. Interestingly, there are some parameter regimes where both $\langle IPR \rangle > 0$ and $\langle NPR \rangle > 0$. This phase is called the mixed or intermediate phase. We note here that the crossover lines $c_1$ and $c_2$ in our phase diagram Fig. \ref{phase_diag} are determined after performing finite-size scaling of $\langle IPR \rangle $ and $\langle NPR \rangle $, as discussed in Appendix \ref{finte_size_scaling}. 

Although $\langle PR \rangle$ gives an estimation of the localization length, one can independently calculate it from the Lyapunov exponent. By performing a transfer matrix calculation, we can calculate the Lyapunov exponent ~\cite{LE_1, LE_2}. For our system, this exponent is defined as follows:
\begin{equation} \label{LE}
    \Gamma(E) = \lim_{N \to \infty} \frac{1}{2N} \ln{\norm{M_N}}, 
\end{equation} 
where $\norm{M_N}$ represents the norm of the matrix $M_N$. This matrix is the product of transfer matrices $M_N =\prod_{k=1}^{N} T_k$. In our problem, the transfer matrix $T_k$ is a $4 \cross 4$ matrix, 
\begin{equation}
T_k =
\begin{pmatrix}
-\dfrac{E}{t_A} & -\dfrac{t'}{t_A} & -1 & 0 \\[10pt]
-\dfrac{t' e^{-i\theta_k}}{t_B} & -\dfrac{E e^{-i\theta_k}}{t_B} & 0 & -\dfrac{e^{-i(\theta_{k-1}+\theta_k)}}{t_B} \\[10pt]
1 & 0 & 0 & 0 \\[10pt]
0 & 1 & 0 & 0
\end{pmatrix}.
\end{equation}
For each eigenenergy $E_i$, we first numerically calculate $\Gamma(E_k)$. From this result we estimate average localization length in the following way:
\begin{equation}
    \overline{\xi}=\frac{1}{2N}\sum_{k=1}^{2N}\frac{1}{\Gamma (E_k)}.
\label{Eq: xi_lyp}
\end{equation}

\subsection{Dynamics tools} \label{dyna_tools}
We also study single-particle dynamics to analyze the localization behavior of our system. 
We initialize the system by placing one particle at the center of the leg $A$, and then let it evolve under the total Hamiltonian $H$ as 
\begin{equation}
|\psi(t)\rangle = e^{-iHt}|\psi_0\rangle,
\end{equation}
where the initial state is given by $|\psi_0\rangle = |\frac{N}{2}\rangle$. The normalized time-evolved state projected onto leg $A$ is given by:
\begin{equation}
|\psi_A(t)\rangle = \frac{\hat{P}_A|\psi(t)\rangle}{\|\hat{P}_A|\psi(t)\rangle\|},
\end{equation}
where $\hat{P}_A=\sum_{m=1}^N |m\rangle \langle m|$. The spread of the particle along the leg $A$ can be measured from its mean squared displacements (MSD). This quantity, which is the variance of the probability distribution associated with the projected state $|\psi_A(t)\rangle$, is given by
\begin{equation}
    \sigma^2(t)= \sum_{m=1}^N m^2 |\psi_A(m,t)|^2  - \left[\sum_{m=1}^{N} m |\psi_A(m,t)|^2 \right]^2.
    \label{variance}
\end{equation}
Typically, in the long-time limit, the MSD scales as follows\cite{exponent_sambudhhya}:
\begin{equation} \label{msd_scaling}
    \sigma^2(t) \sim t^{\gamma},
\end{equation}
with $0\le \gamma \le 2$. In a delocalized phase, the ballistic transport is expected with $\gamma = 2$. On the other hand, $1 < \gamma < 2$ ($0 < \gamma < 1$) indicates superdiffusive
(subdiffusive) behavior in the dynamics. The case $\gamma = 1$ represents the diffusive transport, and in the localized phase one finds $\gamma = 0$. 

The saturation value of MSD also plays an important role in understanding localization behavior of a system. The saturation value is given by
\begin{equation}
    \sigma^2_{\text{sat}} = \lim_{T \to \infty} \frac{1}{T-T_0} \int_{T_0}^{T}\sigma^2(t)\, dt,
    \label{sigma_sat}
\end{equation}
where $T_0$ is the initial time, which is normally taken to be 0. This quantity scales as 
\begin{equation}\label{msd_nu}
    \sigma^2_{sat}\sim N^{\nu}
\end{equation}
with $0\le \nu \le 2$. A completely localized (delocalized) phase corresponds to $\nu = 0$ (2). 

\section{Numerical analysis}
\label{sec: numerical analysis}
\begin{figure}[t] 
    \centering
\includegraphics[width=0.44\textwidth]{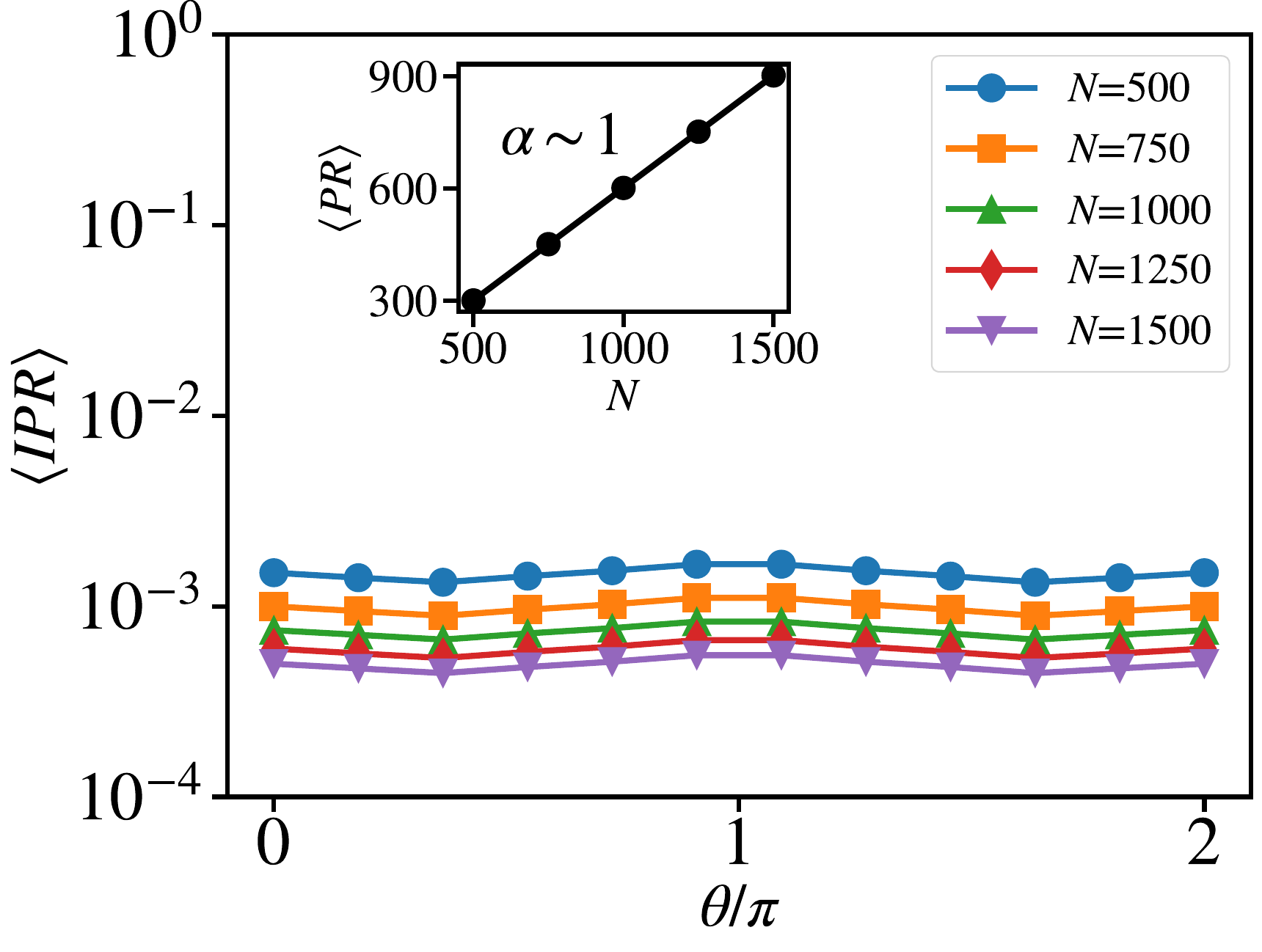}
\caption{Uniform magnetic flux: Average IPR vs. $\frac {\theta}{\pi}$ plot for $N= 500, 750, 1000, 1200 ~\mathrm{and}~ 1500$ respectively. The inset shows the variation of average PR with $N$ (for $\theta=\frac{\pi}{4}$).}
\label{uni_stat}
\end{figure}

\begin{figure}[t]
    \centering
   \includegraphics[width=0.44\textwidth]{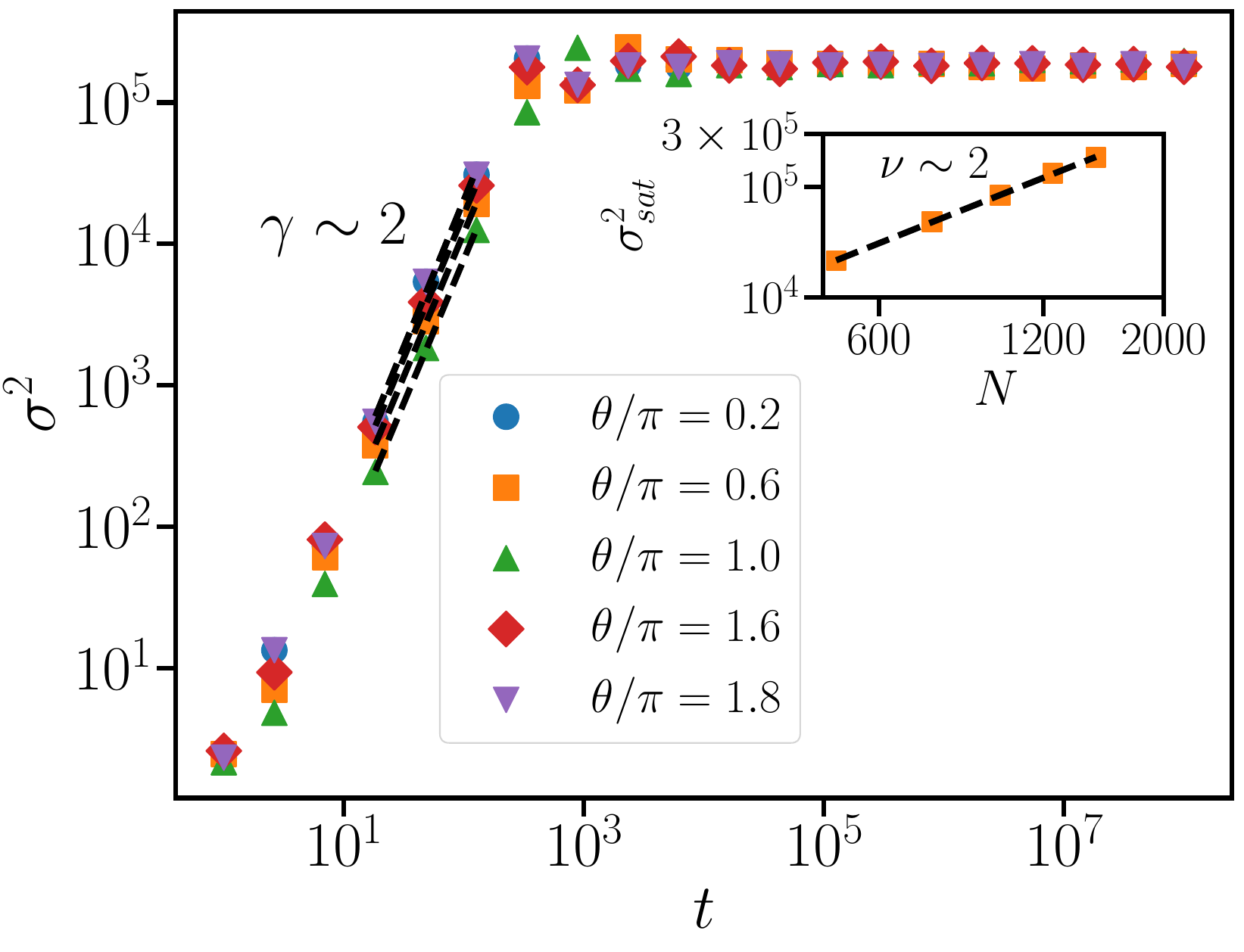}
    \caption{Uniform magnetic flux: $\sigma^2$ vs. $t$ plot for different $\theta$ values for $N=1500$. The black dashed lines show the fitting to the functional form $\sigma^2\sim t^\gamma$. Inset: $\sigma^2_{sat}$ vs. $N$ plot. Black dashed line in the inset shows the fitting to the functional form $\sigma^2_{sat} \sim N^\nu$.}
    \label{dy_uniform}
\end{figure}

\subsection{Uniform magnetic flux} 
First, we discuss the scenario where the magnetic flux is uniform across the lattice. This corresponds to a constant Peierls phase $\theta$ (Eq. \ref{theta_n} with $V=0$). For different values of $\theta$, we calculate the average inverse participation ratio $\langle IPR \rangle$, as defined in Eq. \ref{ipr_av}, to analyze the overall localization properties of the system under the uniform magnetic field. The results are presented in Fig. \ref{uni_stat} for the subsystem sizes $N= 500, 750, 1000, 1250 ~\mathrm{and}~ 1500$. The results show that $\langle IPR \rangle$ remains nearly constant throughout the range of $\theta$. Here we find $\langle IPR \rangle \sim \frac{1}{2N}$ - this scaling indicates that the eigenstates are extended in nature. In the inset, we also plot the average participation ratio $\langle PR \rangle$ (see Eq.~\ref{pr_av}), as a function of $N$. The linear increase of $\langle PR \rangle$ with $N$ (with an exponent $\alpha \sim 1$) further confirms the extended nature of the eigenstates. 

In a similar spirit, we also study the dynamical properties of the system where the magnetic flux is uniform throughout the system. We initiate the process ($t=T_0=0$) by placing a particle in the middle of the leg $A$. In Fig.~\ref{dy_uniform}, we plot the variance $\sigma^2$ (see Eq.~\eqref{variance}) of the distribution as a function of time $t$ for different values of $\theta$. Before saturation, the variance is seen to follow the scaling relation $\sigma^2 \sim t^\gamma$. We find that the exponent $\gamma=1.99$ for different values of $\theta$. This value of $\gamma$ suggests a ballistic spreading of the wavefunction with the time along the lattice.
Furthermore, in the inset we present the long-time averaged saturation values of the variance ($\sigma^2_{\text{sat}}$; see Eq.~\eqref{sigma_sat})  as a function of $N$, and see that it increases with $N$. This result is consistent with the fact that the states are delocalized. 

Both static and dynamic results collectively demonstrate that a uniform magnetic flux does not induce localization in the ladder system. 

\begin{figure}[t]
    \centering
\includegraphics[width=0.44\textwidth]{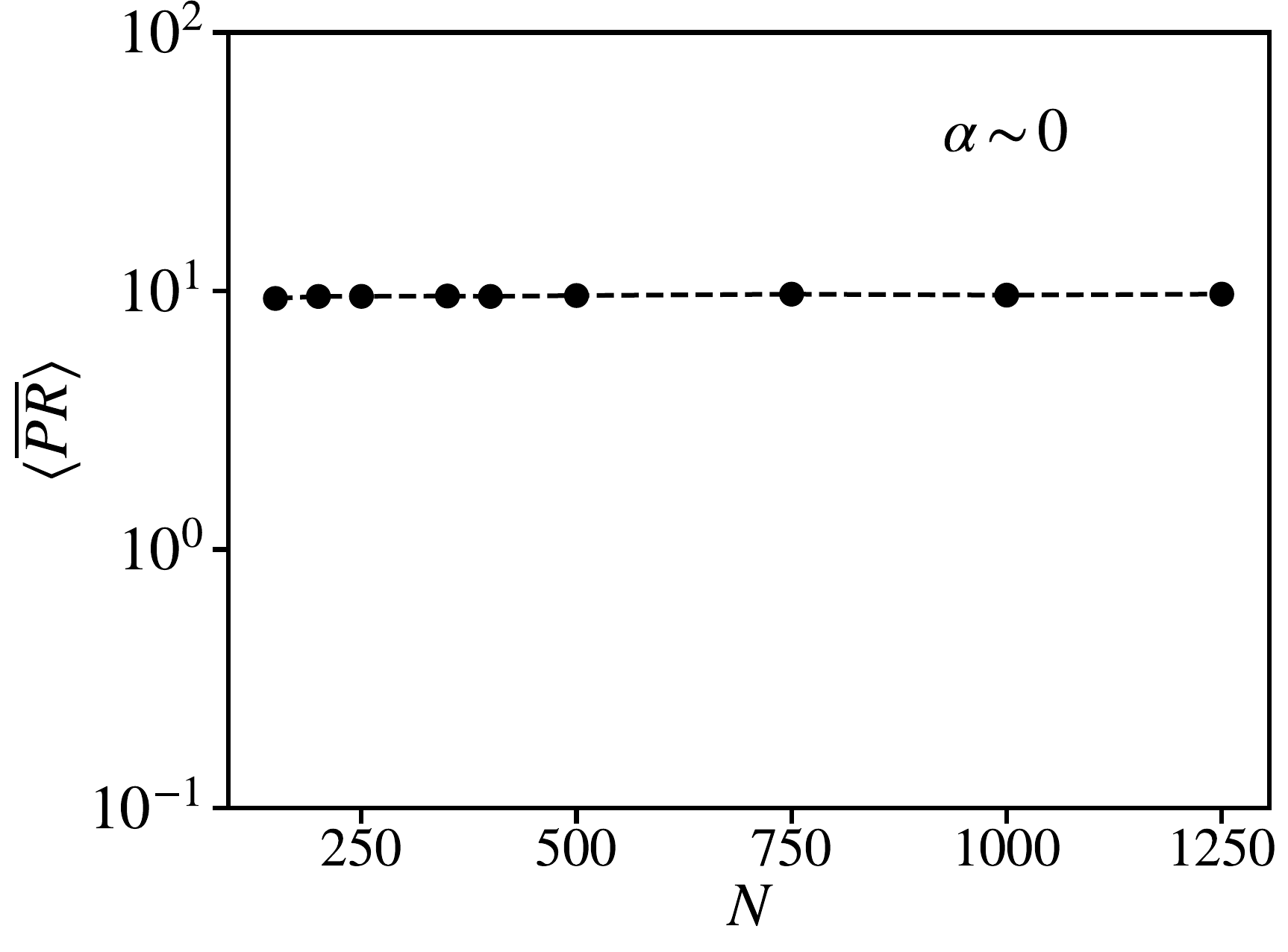}
\caption{Random magnetic flux: Average PR vs. $N$ plot, where average is performed over eigenstates and over different random realizations of the Peierls phase $\theta_n \in [0, 2\pi)$.}
\label{rndm_stat}
\end{figure}

\begin{figure}[t]
    \centering
  \includegraphics[width=0.44\textwidth]{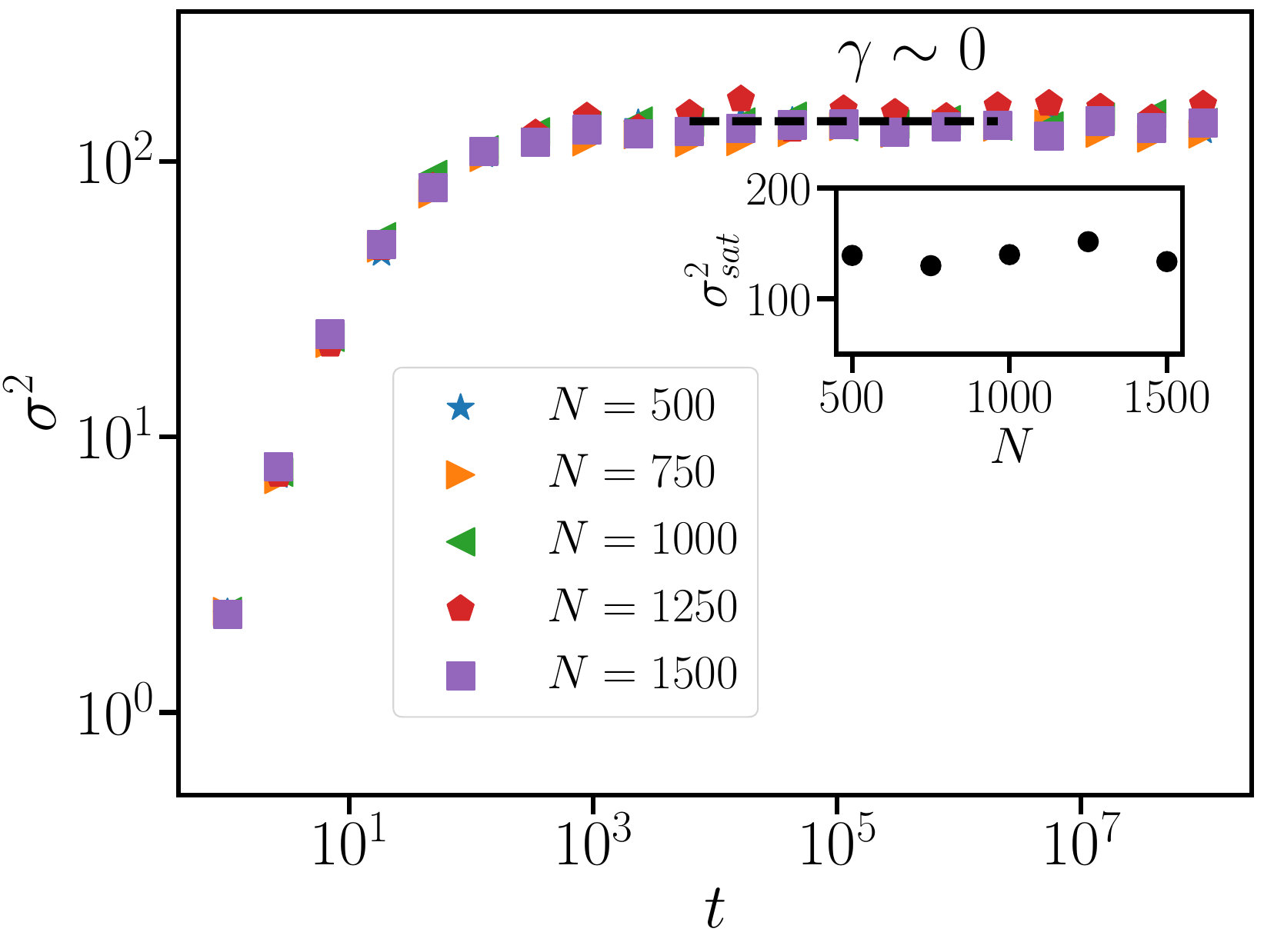}
      \caption{Random magnetic flux: $\sigma^2$ vs. $t$ plot for random $\theta_n \in [0, 2\pi)$ for  different $N$. The black dashed lines show the fitting to the functional form $\sigma^2\sim t^\gamma$. Inset: $\sigma^2_{sat}$ vs. $N$ plot.}
    \label{dy_random}
\end{figure}

\subsection{Random magnetic flux}\label{ran_mag}
In this section, we consider the scenario where we have random magnetic flux across the system. 
In this case, the Peierls phase $\theta_n$, for each $n$, is taken randomly from a uniform distribution $\theta_n \in [0,2\pi)$. For a given realization of the Peierls phases, we first calculate the state averaged participation ratio $\langle PR \rangle$ (see Eq.~\ref{pr_av}).  
Subsequently, we average $\langle PR \rangle$ over at least 20 independent random realizations of Peierls phases to obtain the disorder-averaged quantity $\langle \overline{PR} \rangle$. The finite-size scaling of $\langle \overline{PR} \rangle$, shown in Fig.~\ref{rndm_stat}, indicates that the average PR remains nearly constant with increasing system size ($\alpha \sim 0$). This behavior suggests that the eigenstates become localized with a finite localization length.

For the random flux case, we also study the dynamical properties of the system to verify the signature of localization. 
From the results presented in Fig.~\ref{dy_random}, we see that although $\sigma^2$ initially grows with time, it saturates to a comparatively small saturation value in some intermediate time, and fitting to the function $\sigma^2\sim t^\gamma$ yields $\gamma\sim0$. Also, in the inset, it is seen that the long-time averaged saturation value ($\sigma^2_{\text{sat}}$; see Eq.~\eqref{sigma_sat}) remains constant with increasing system size $N$. 
These findings suggest that the random magnetic flux localizes the states, which gives no transport through the lattice in the thermodynamic limit.   

\subsection{Quasiperiodic magnetic flux} 
We have performed a detailed analysis using both static and dynamical measures (as discussed in Sec.~\ref{diagn_tools}) to construct the schematic phase diagram shown in Fig.~\ref{phase_diag}. In this section, we present the numerical results along the $V=1$ and $V=3$ lines. Since the $V=1$ line encompasses all three phases — completely delocalized, mixed and localized — we perform an even more extensive analysis to identify the phase boundaries accurately for the $V=1$ line. 
The analysis corresponding to the $\lambda=0$ and $\lambda \to 1$ limits is provided in Appendix~\ref{lambda_lines}.

\subsubsection{$V=1$ and $0\le \lambda < 1$}
\begin{figure}[t]
    \centering
\includegraphics[width=0.48\textwidth]{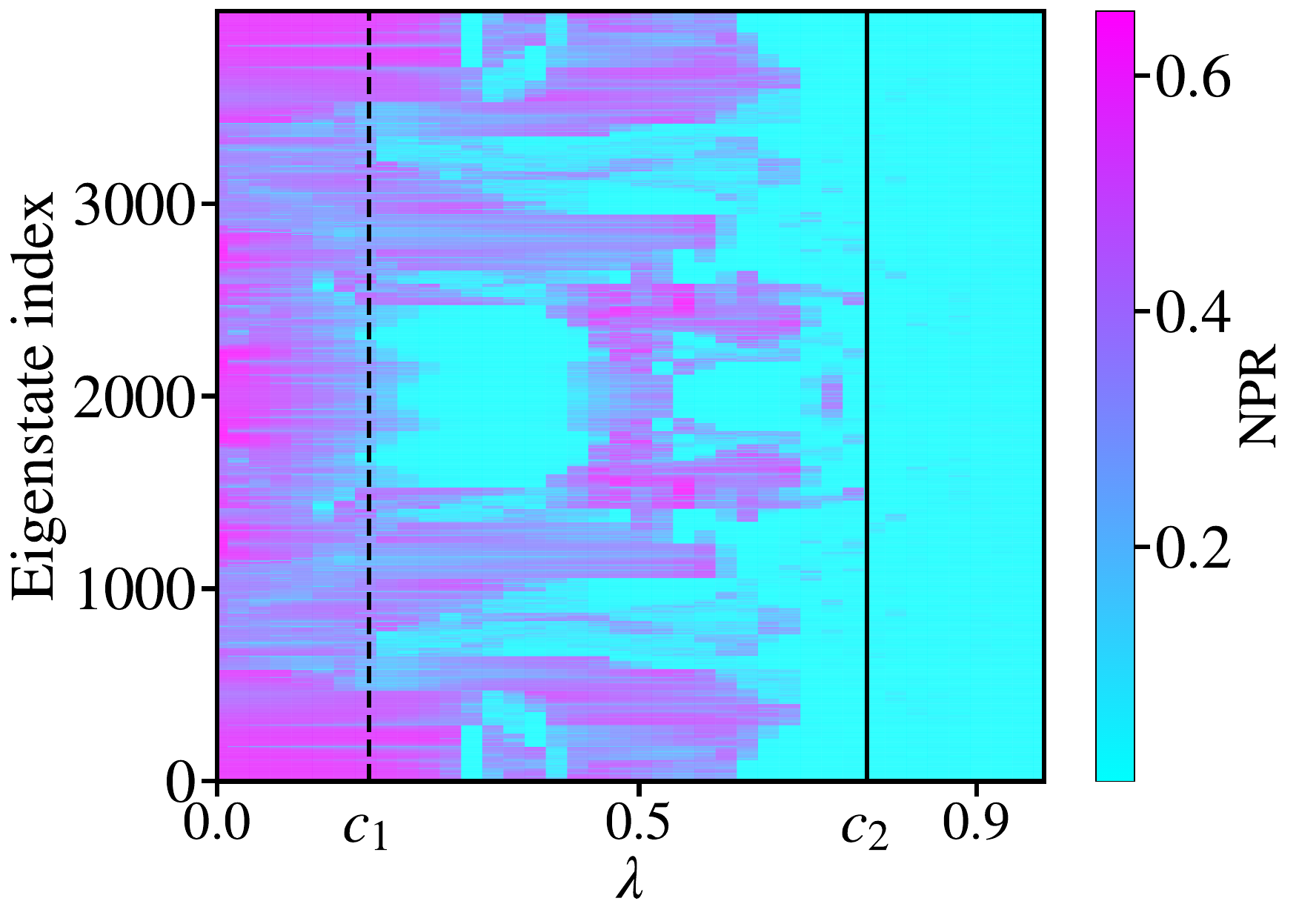}
\caption{Quasiperiodic magnetic flux ($V=1$): Contour plot of $NPR$ as a function of $\lambda$ and eigenstate index for $N=2000$. }
\label{cont_plt}
\end{figure}

\begin{figure}[t]
    \centering
\includegraphics[width=0.44\textwidth]{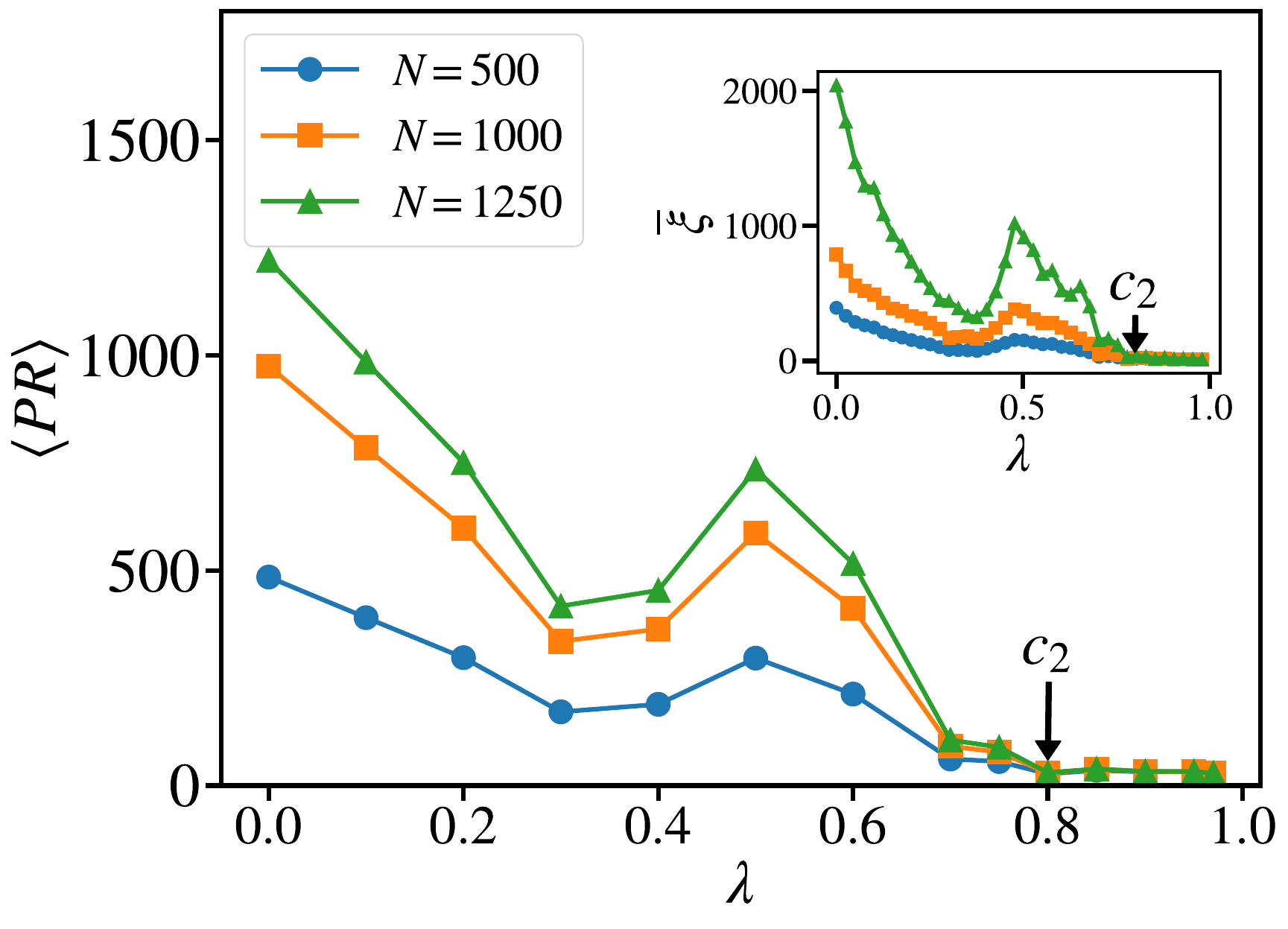}
\caption{Quasiperiodic magnetic flux ($V=1$): Average PR vs. $\lambda$ plot for $N=500, 1000, 1250$. Inset: $\overline{\xi}$ vs. $\lambda$ plot.}
\label{pr vs L V1}
\end{figure}

\begin{figure}[t]
    \centering
\includegraphics[width=0.44\textwidth]{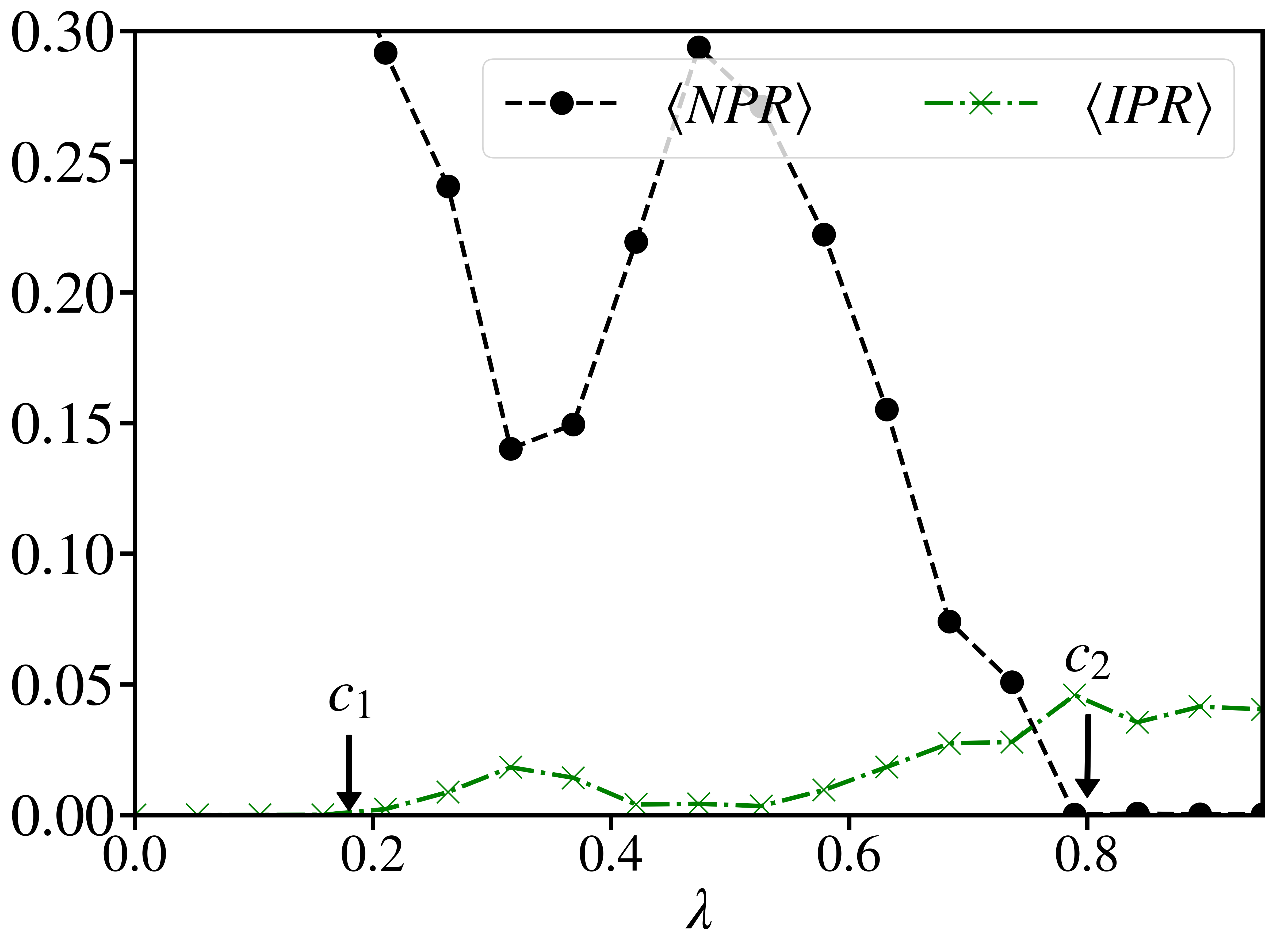}
\caption{Quasiperiodic magnetic flux ($V=1$): Plot of average $IPR$ and $NPR$ as a function of $\lambda$ in the thermodynamic limit ($N \to \infty$).}
\label{NPR_IPR_V_1}
\end{figure}

\begin{figure*}[t]
    \centering
\includegraphics[width=1\textwidth]{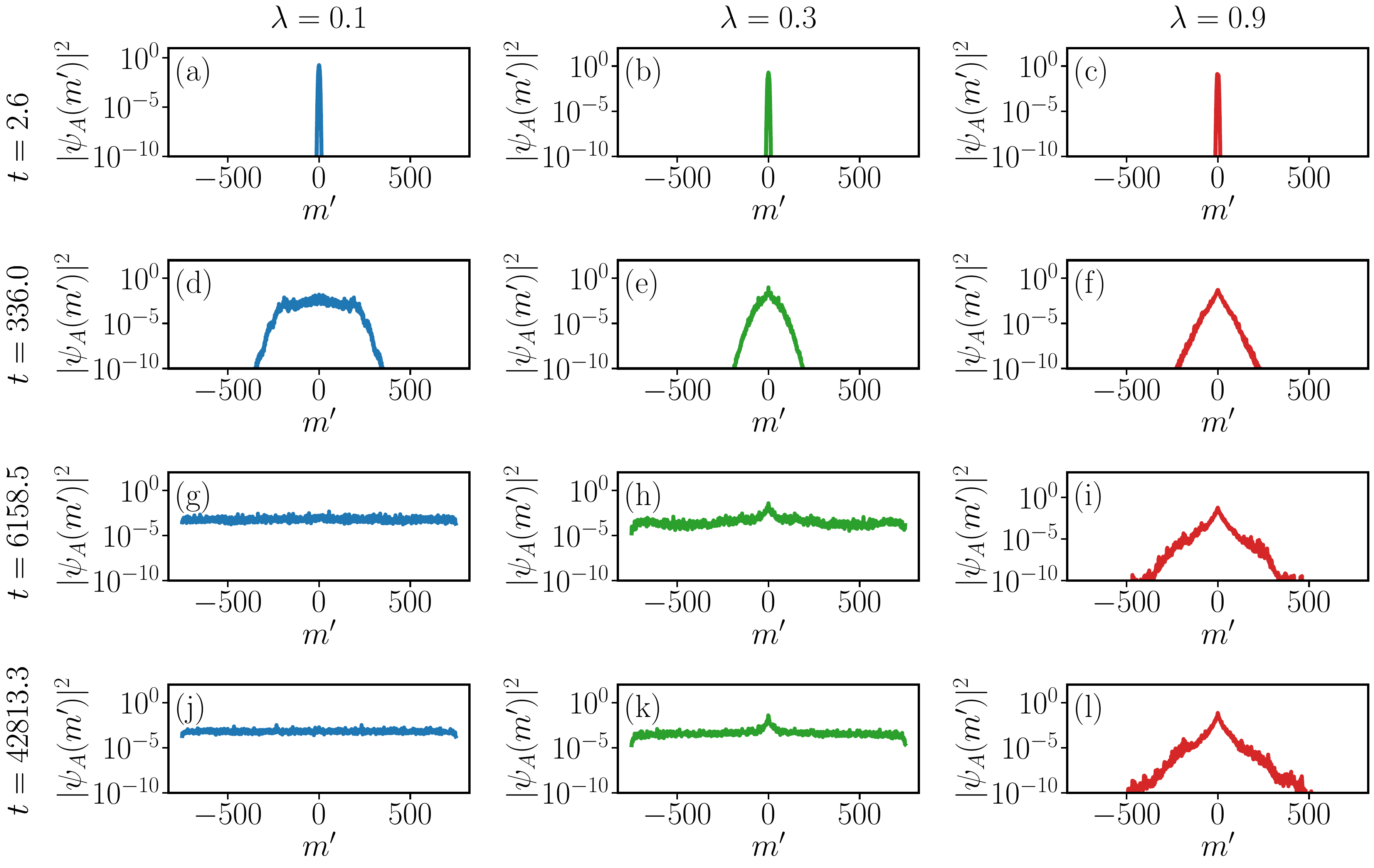}
\caption{Quasiperiodic magnetic flux ($V=1$): Probability distribution $|\psi_A(m')|^2$ vs. $m'$ plot for different $\lambda$ and different instant of time $t$, for  $N=1500$. The particle is initiated from the middle of the leg $A$, i.e., $m=N/2$, and here $m'=m-N/2$.}
\label{dy_prop}
\end{figure*}

We first focus on the eigenstate properties of the system and investigate how $PR$ behaves as a function of $\lambda$ for $V=1$. We compute $NPR$ for individual eigenstates and present the results in Fig.~\ref{cont_plt} as a contour plot for a fixed system size $N=2000$. 
We expect that low (high) values of $NPR$ correspond to states that are more localized (delocalized) in nature. We find that for $\lambda \gtrsim 0.8$ (indicated by the solid vertical line in the plot), the $NPR$ values for all eigenstates are relatively small, signifying a localized phase. In contrast, for $\lambda \lesssim 0.8$, the $NPR$ values for at least some states are comparatively large, indicating existence of extended states. 
Moreover, we observe that for $\lambda \gtrsim 0.18$ (indicated by a vertical dashed line), mostly there exist pockets of eigenstates within the spectrum that retain relatively low $NPR$ values, suggesting the coexistence of localized  states within an otherwise delocalized background.

However, the value of $NPR$ alone cannot determine whether the system is overall localized or delocalized. Hence, in Fig.~\ref{pr vs L V1}, we study the finite-size scaling of $PR$ averaged over all eigenstates. We find that for $\lambda \gtrsim 0.8$, $\langle PR \rangle$ does not scale with $N$, signifying an overall localized phase. In contrast, for smaller values of $\lambda$, $\langle PR \rangle$ increases with $N$, indicating an overall delocalized phase. These results are consistent with our previous eigenstate-resolved $NPR$ analysis. It also suggests that the presence of pockets of localized states in the range $0.18 \lesssim \lambda \lesssim 0.8$ does not destroy the overall delocalized nature of the phase, as inferred from the finite-size scaling of $PR$. 

We further validate our findings by computing the Lyapunov exponent and the corresponding average localization length $\overline{\xi}$ (see Eq.~\eqref{Eq: xi_lyp}), as shown in the inset of Fig.~\ref{pr vs L V1}. A more detailed study of the Lyapunov exponent can be found in Appendix \ref{lya_expo}. These results are in complete agreement with the behavior observed in our study of $\langle PR \rangle$.

Now, from the previous analyses based on the eigenstate-resolved $NPR$, $\langle PR \rangle$, and $\overline{\xi}$, we are able to identify the point on the $c_2$ line corresponding to $V=1$. Our estimation shows that $\lambda \simeq 0.8$ for the point, above which the system is completely localized. 

From the eigenstate-resolved $NPR$ plot, we also observe that for $\lambda \lesssim 0.18$, almost all eigenstates appear to be delocalized. In contrast, for $0.18 \lesssim \lambda \lesssim 0.8$, there is a coexistence of states with both high and low $NPR$, suggesting a mixed character of the spectrum.
In Fig.~\ref{NPR_IPR_V_1}, we show the variation of $\langle NPR \rangle$ and $\langle IPR \rangle$, extrapolated to the thermodynamic limit ($N \to \infty$). The extrapolation procedure is discussed in Appendix~\ref{finte_size_scaling}. The figure shows that for $0 \le \lambda \lesssim 0.18$, $\langle IPR \rangle$ vanishes while $\langle NPR \rangle$ remains finite, indicating a completely delocalized phase. 
For $0.18 \lesssim \lambda \lesssim 0.8$, both $\langle IPR \rangle$ and $\langle NPR \rangle$ are finite, signifying an intermediate (mixed) phase. On the other hand, in the region $0.8 \lesssim \lambda < 1$, $\langle NPR \rangle$ vanishes while $\langle IPR \rangle$ remains finite, identifying a completely localized phase, consistent with our earlier observations.

\begin{figure}[t]
    \centering
\includegraphics[width=0.44\textwidth]{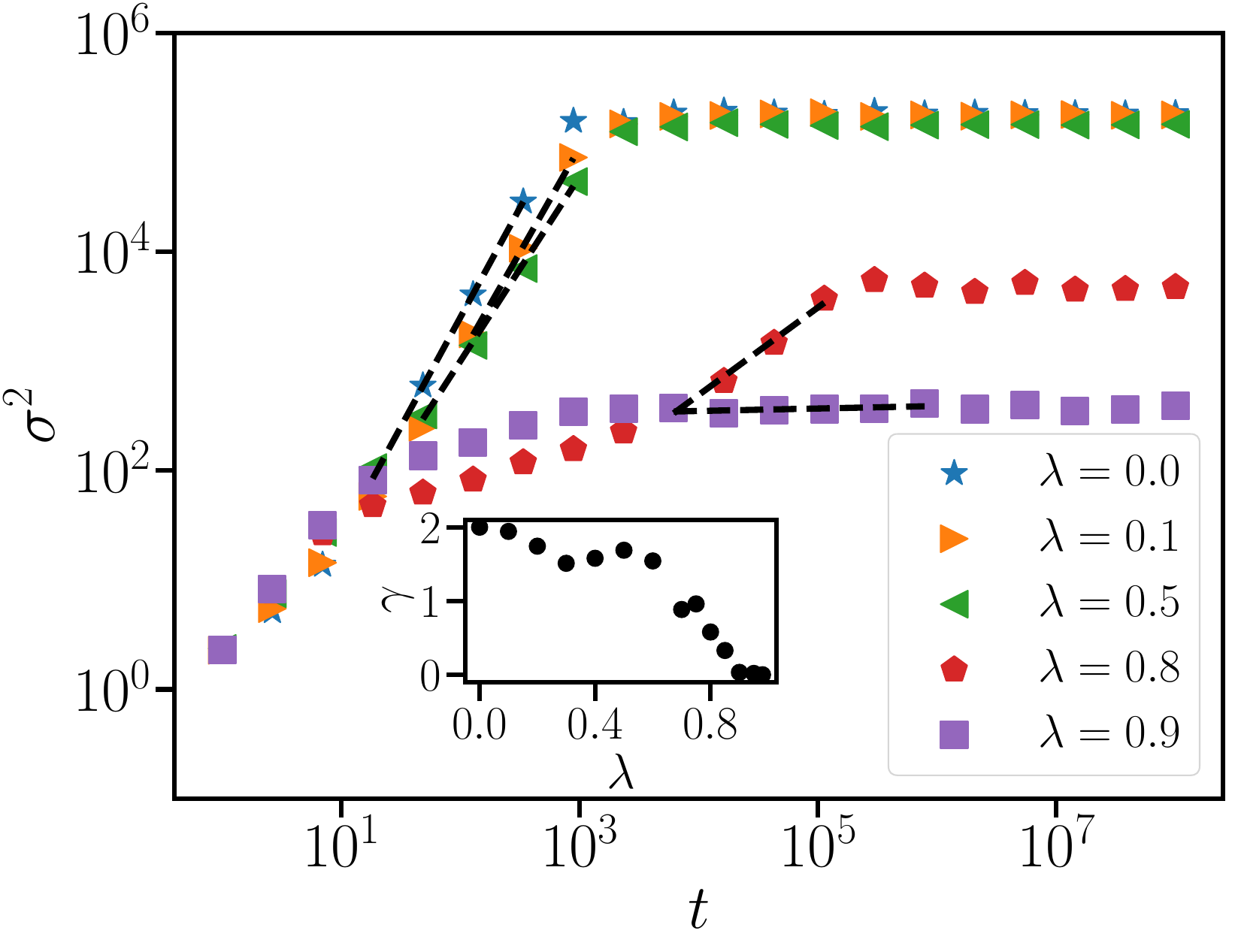}
 \caption{Quasiperiodic magnetic flux ($V=1$): $\sigma^2$ vs $t$ plot for different $\lambda$ values for $N=1500$. The black dashed lines show the fitting to the functional form $\sigma^2\sim t^\gamma$. Inset: $\gamma$ vs. $\lambda$ plot.}
\label{dy_sigma_GAA}
\end{figure}

\begin{figure}[t]
    \centering
\includegraphics[width=0.44\textwidth]{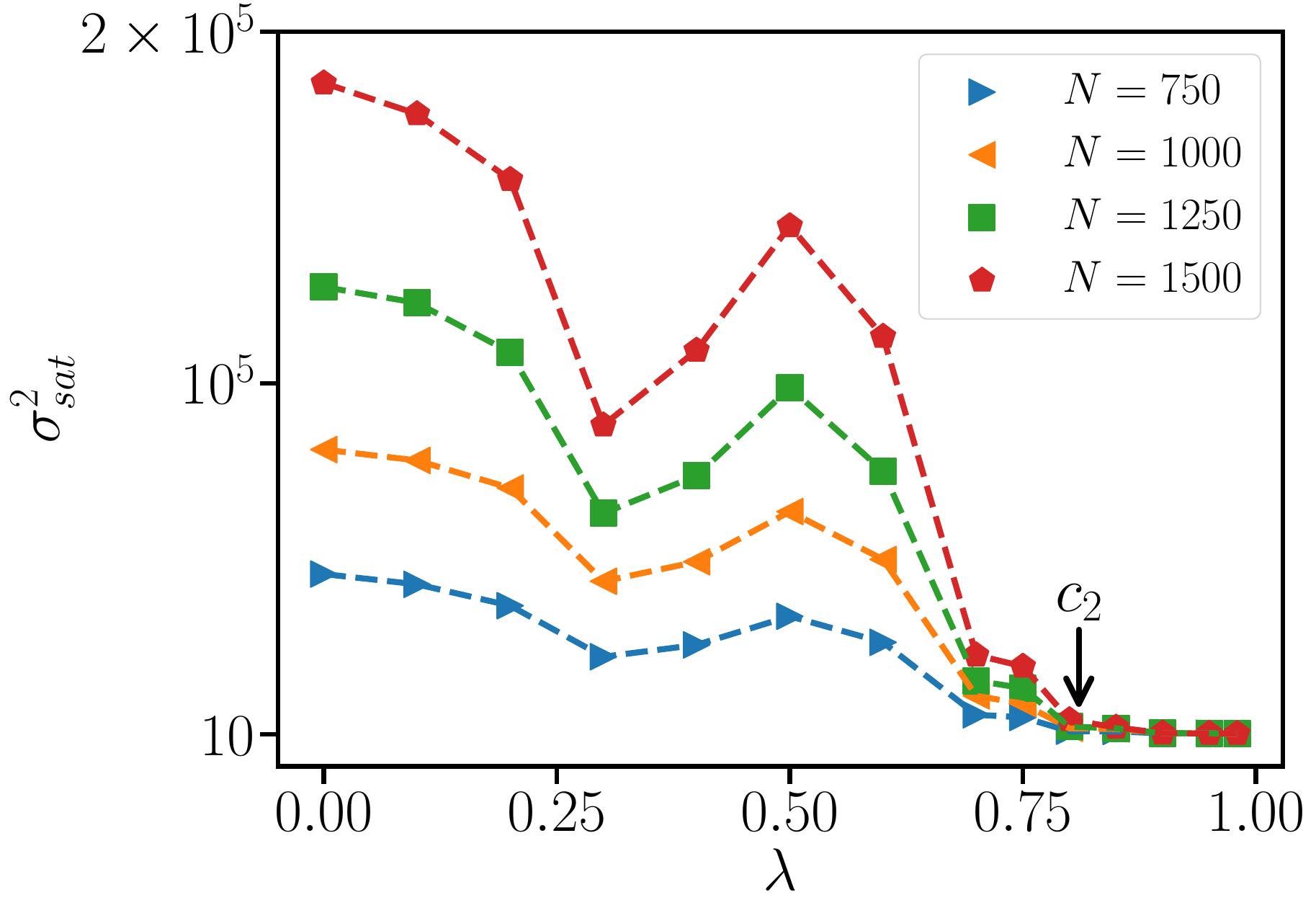}
\caption{Quasiperiodic magnetic flux ($V=1$): $\sigma^2_{sat}$ vs. $\lambda$ plots for different $N$ values.}
\label{dy_sigma_sat_GAA}
\end{figure}

In the intermediate mixed-phase region, we also observe a reentrant localization phenomenon (see Figs. \ref{cont_plt}, \ref{pr vs L V1}, and \ref{NPR_IPR_V_1}). As $\lambda$ increases, the system first evolves from a delocalized phase to a more localized phase, then reenters a more delocalized phase, and finally undergoes a transition to a completely localized phase for larger values of $\lambda$. This reentrant behavior can be qualitatively understood in terms of the non-monotonic evolution of the spatial correlations of the Peierls phases with the parameter $\lambda$. A detailed discussion is provided in Appendix \ref{appn_reentrant}.

Next, we study the single-particle dynamics. As discussed in Sec.~\ref{diagn_tools}, we first investigate the time evolution of the projected wave-packet dynamics $|\psi_A|^2$. Figure~\ref{dy_prop} shows snapshots of $|\psi_A|^2$ at four different times for some representative values of $\lambda$. 
As expected, for $\lambda = 0.1$ (completely delocalized phase), $|\psi_A|^2$ spreads rapidly across the $A$ leg of the ladder with increasing time. For $\lambda = 0.3$ (intermediate phase), the wave-packet propagation is qualitatively similar; however, even at long times (when the wave packet has already reached the edges of the $A$ leg), a small peak persists at the initial position. This indicates a residual memory of the initial state (see Appendix~\ref{memory} for a detailed analysis of this memory effect). 
On the other hand, for $\lambda = 0.9$ (completely localized phase), the wave packet spreads only for a short initial time and then ceases to propagate further. Consequently, for sufficiently large $N$, it never reaches the boundaries, and the dynamics effectively freezes, which is a characteristic signature of the localized phase.

Next, to quantify the spread of the wave packet, we compute the MSD, $\sigma^2$, and study its time evolution, as shown in Fig.~\ref{dy_sigma_GAA}. As expected, $\sigma^2$ increases with time and eventually saturates. To further characterize the growth rate, we plot the dynamical exponent $\gamma$ in the inset. To reliably extract the dynamical exponent $\gamma$ for the delocalized or mixed regime, we have identified a fitting time window that is both sufficiently long to capture the asymptotic scaling behavior and unaffected by finite-size saturation. Since the saturation time depends on the Hamiltonian parameters, the appropriate fitting window must be determined separately for each set of parameters. In our analysis, we systematically increase the system size and verify that the fitted value of $\gamma$ remains unchanged within the chosen pre-saturation time window. Moreover, to strengthen our findings, in the Appendix.~\ref{appndx_finite_time}, we have also incorporated an independent approach to extract the dynamical exponent $\gamma$ based on finite-time scaling analysis. We find that in the completely delocalized phase ($0 \le \lambda \lesssim 0.18$), the dynamics is nearly ballistic ($\gamma \simeq 2$). In the mixed phase ($0.18 \lesssim \lambda \lesssim 0.8$), the dynamics are sub-ballistic, with $0 < \gamma < 2$. Finally, in the completely localized phase ($\lambda \gtrsim 0.8$), $\gamma \simeq 0$.
Figure~\ref{dy_sigma_sat_GAA} also shows the variation of the saturation value of $\sigma^2$ with system size $N$. As one can already predict, for $\lambda \gtrsim 0.8$, $\sigma^2_{\text{sat}}$ does not scale with $N$, indicating a completely localized phase. In contrast, in both the completely delocalized and mixed phases ($0 \le \lambda \lesssim 0.8$), the saturation value increases with $N$.

\subsubsection{$V=3$ and $0\le \lambda < 1$}

\begin{figure}[t]
    \centering
\includegraphics[width=0.44\textwidth]{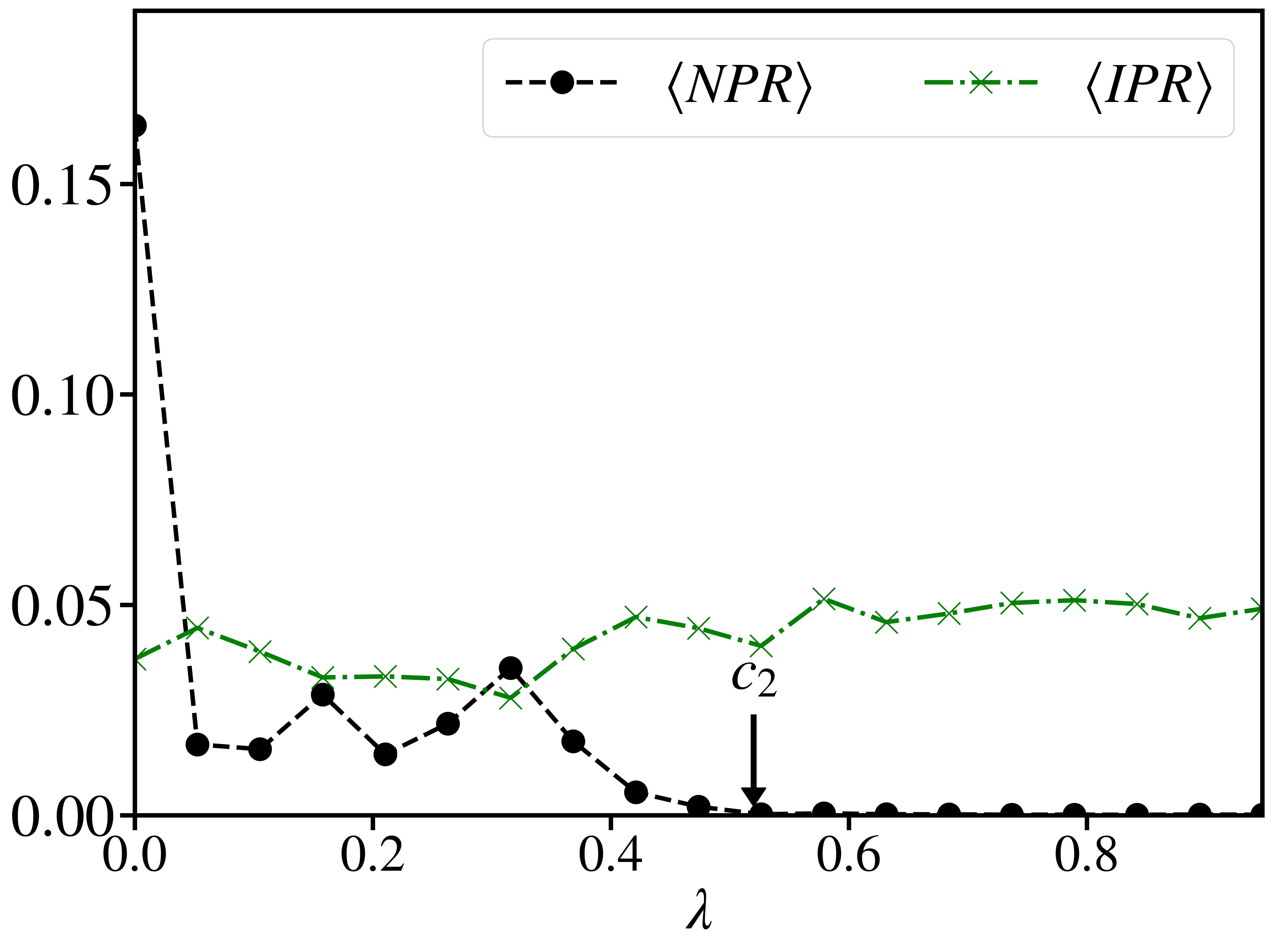}
\caption{Quasiperiodic magnetic flux ($V=3$): Plot of average $IPR$ and $NPR$ as a function of $\lambda$ in the thermodynamic limit ($N \to \infty$).}
\label{Fig:IPR NPR V=3}
\end{figure}

\begin{figure}[t]
    \centering
\includegraphics[width=0.44\textwidth]{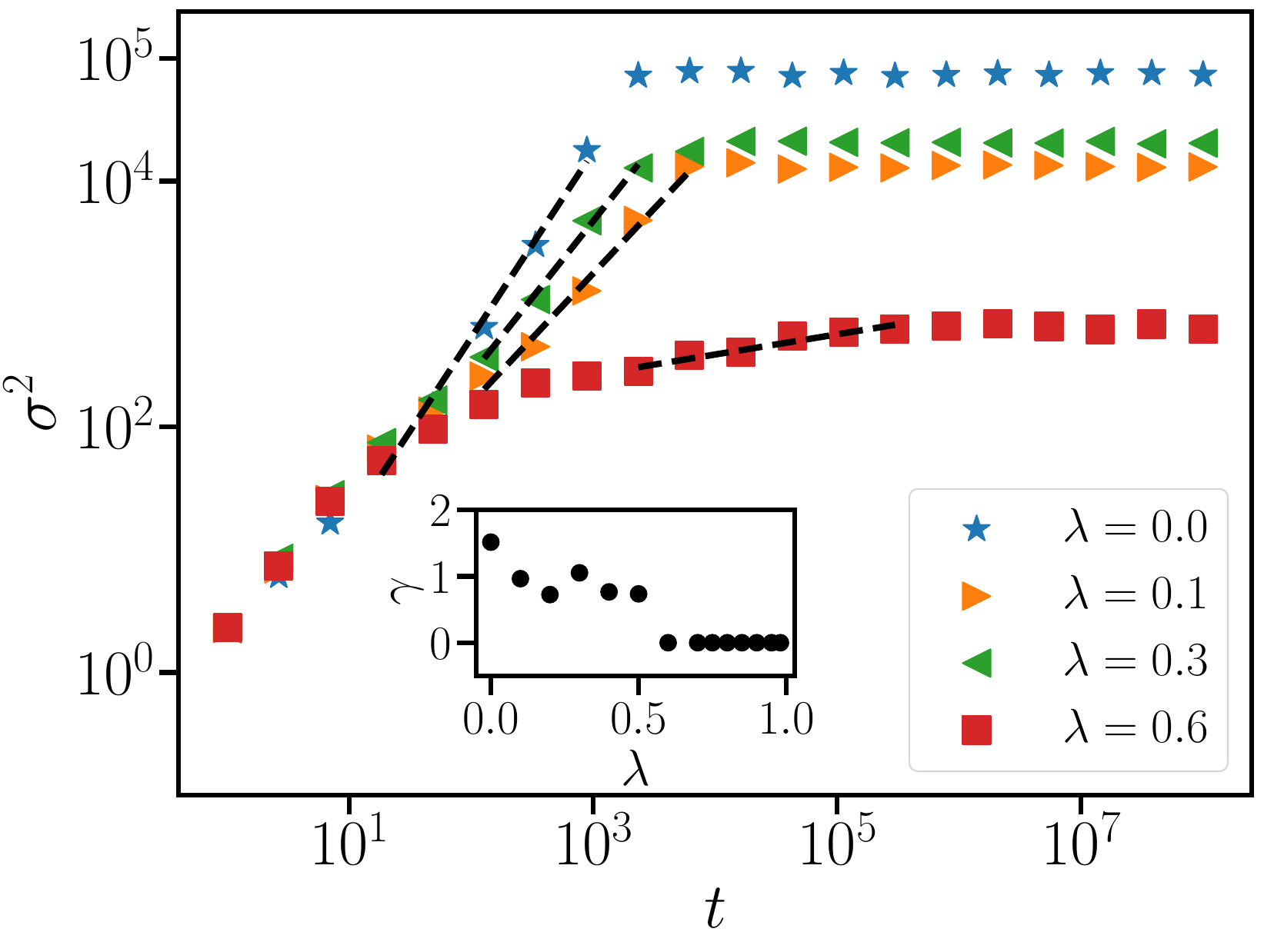}
 \caption{Quasiperiodic magnetic flux ($V=3$): $\sigma^2$ vs $t$ plot for different $\lambda$ values for $N=1500$. The black dashed lines show the fitting to the functional form $\sigma^2\sim t^\gamma$. Inset: $\gamma$ vs. $\lambda$ plot.}
\label{Fig: sigma vs t V=3}
\end{figure}
\begin{figure}[t]
    \centering
\includegraphics[width=0.44\textwidth]{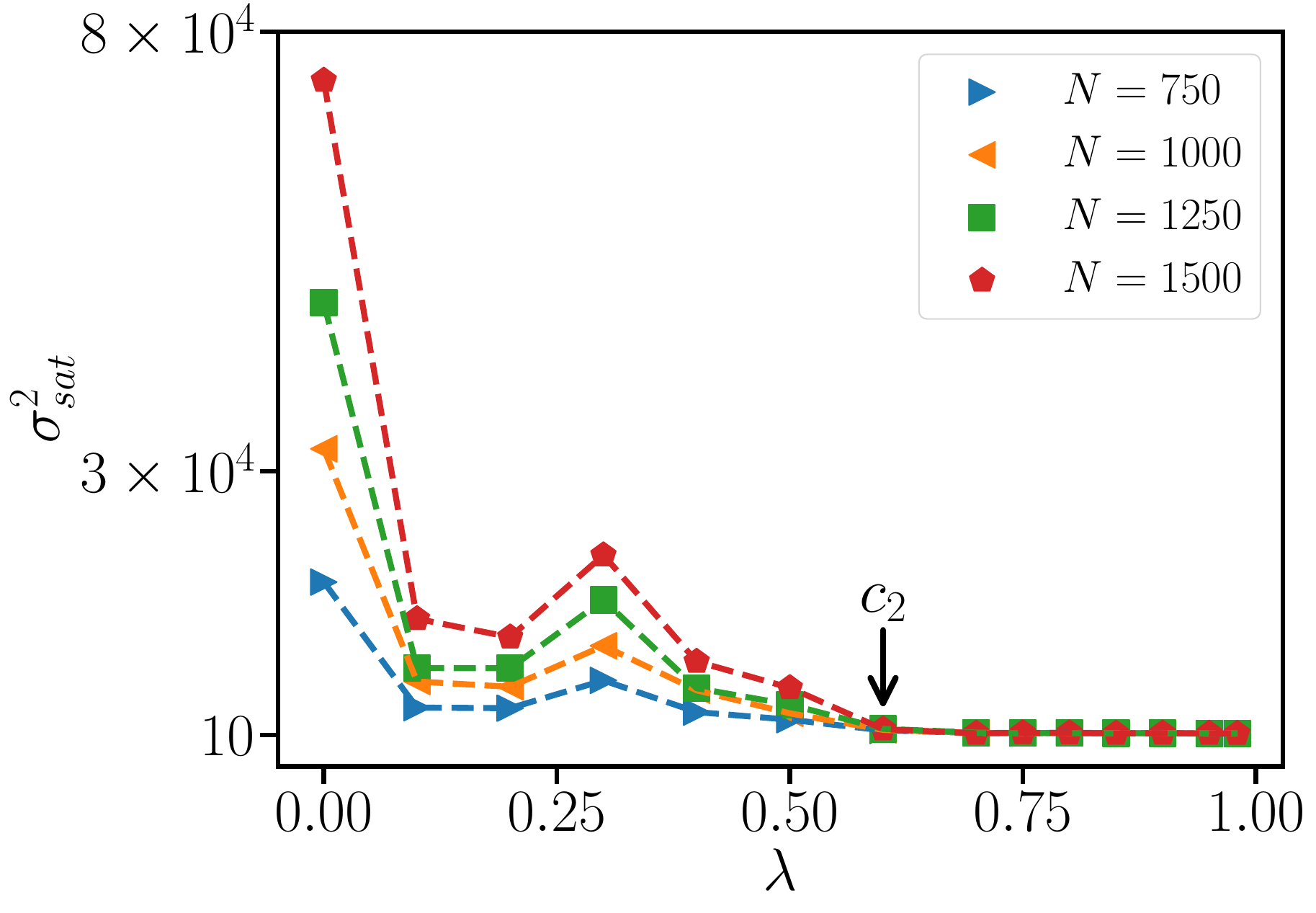}
\caption{Quasiperiodic magnetic flux ($V=3$): $\sigma^2_{sat}$ vs. $\lambda$ plots for different $N$ values.}
\label{Fig: sigma_sat V=3}
\end{figure}

For the $V=3$ line, we again begin by using static measures to identify the different phases. Figure~\ref{Fig:IPR NPR V=3} shows the variation of the averaged IPR and NPR over all eigenstates as a function of $\lambda$, extrapolated to the thermodynamic limit ($N \to \infty$). We find that beyond a critical value $\lambda \simeq 0.55$, $\langle NPR \rangle$ vanishes, while $\langle IPR \rangle$ remains finite, indicating a fully localized phase. Therefore, $\lambda \simeq 0.55$ defines the $c_2$ line in the phase diagram shown in Fig.~\ref{phase_diag}.
On the other hand, for $\lambda \lesssim 0.55$, both $\langle NPR \rangle$ and $\langle IPR \rangle$ remain finite in the thermodynamic limit, signifying the presence of an intermediate phase. Notably, we do not observe a completely delocalized phase along the $V=3$ line.

Next, we study the dynamics. Figure~\ref{Fig: sigma vs t V=3} shows the time evolution of the mean square displacement (MSD) $\sigma^2$, from which we extract the dynamical exponent $\gamma$. Consistent with our previous static analysis, for $\lambda \gtrsim 0.55$, $\gamma$ approaches zero, indicating localized states.
On the other hand, in the intermediate phase ($\lambda \lesssim 0.55$), the exponent $\gamma$ remains finite. However, unlike a completely delocalized phase, the dynamics in this regime is not ballistic; instead, it is sub-ballistic with $0 < \gamma < 2$.

We also analyze the saturation value $\sigma^2_{\mathrm{sat}}$ as a function of $\lambda$ for different system sizes $N$, as shown in Fig.~\ref{Fig: sigma_sat V=3}. We find that for $\lambda \gtrsim 0.55$, the saturation value does not scale with $N$, confirming the localized phase. In contrast, in the intermediate phase, $\sigma^2_{\mathrm{sat}}$ increases with $N$, indicating extended behavior.

\section{Semiclassical Analysis}\label{sec: semiclassical analysis}

\begin{figure*}[t]
    \centering
\includegraphics[width=0.32\textwidth]{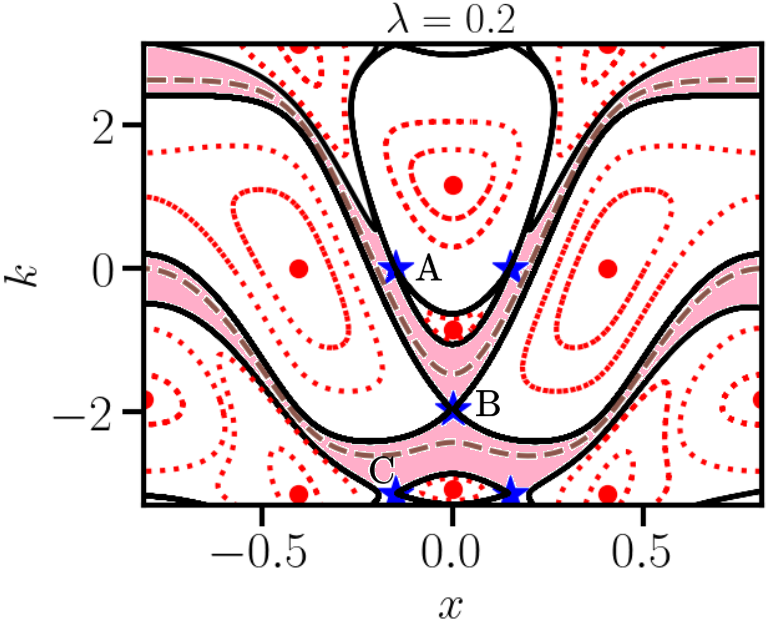}
\includegraphics[width=0.32\textwidth]{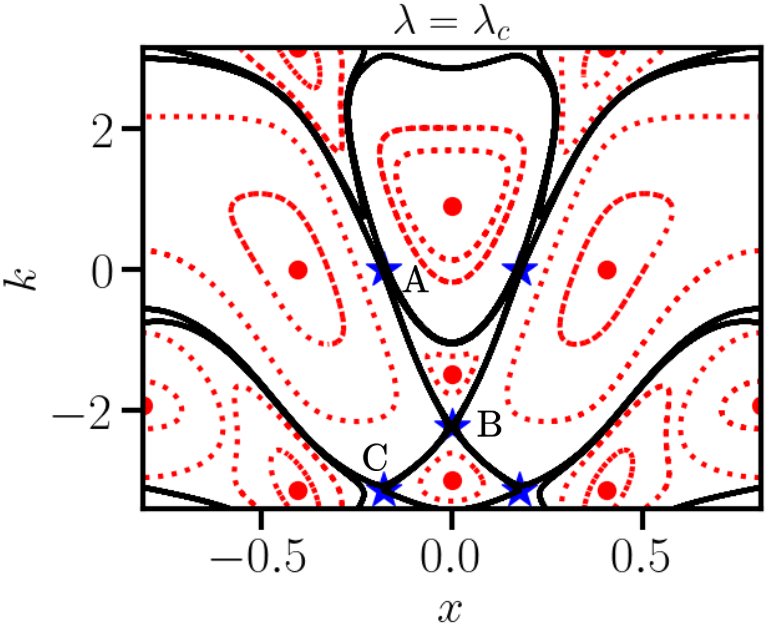}
\includegraphics[width=0.32\textwidth]{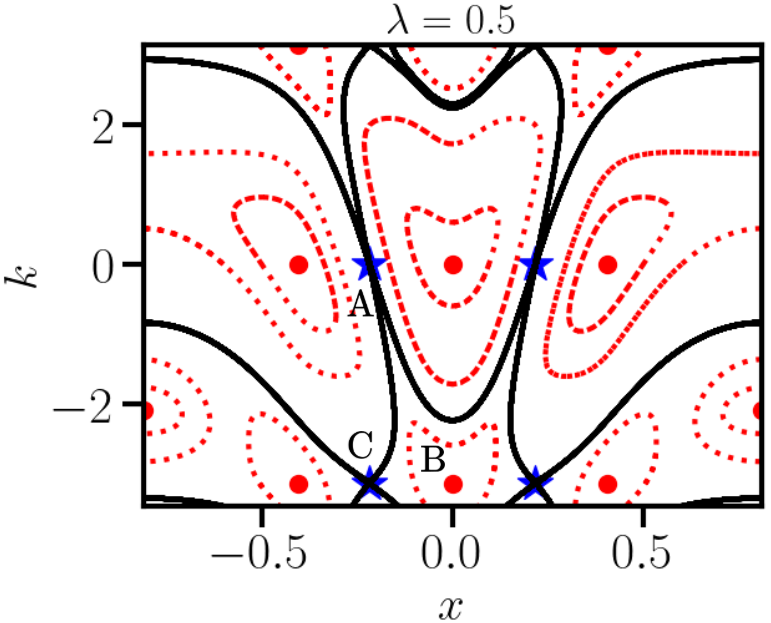}
\caption{Classical trajectory evolution for  magnetic flux $\theta(x)=\frac{\pi \cos(2\pi\beta x)}{1-\lambda \cos(2\pi \beta x)}$, and $\lambda_c\simeq 0.298$. Red circles represent stable fixed points, while blue stars denote saddle points.}
\label{classical_GAA}
\end{figure*}

It has been shown previously that localization--delocalization transitions in lattice models with deterministic quasiperiodic onsite potentials can be captured within a semiclassical framework~\cite{semi_classical_10,chatterjee2026quantum}. In contrast, for systems with true disorder, the situation is markedly different, and similar semiclassical analysis fails to reproduce the corresponding quantum behavior. In our case, the magnetic flux is also deterministic and quasiperiodic in nature and can induce localization--delocalization transitions in the lattice model. This observation motivates us to perform a similar semiclassical analysis, following the approach of Ref.~\cite{chatterjee2026quantum,semiclassical2}.

As a first step, we propose an approximate semiclassical continuum Hamiltonian corresponding to our quantum lattice Hamiltonian $H$ (see Eq.~\eqref{ham_eqn}). 
In the limit when   $\theta_n=\theta$ (uniform magnetic flux), the lattice  Hamiltonian can be diagonalized in the momentum basis by imposing periodic boundary conditions. In the momentum space, the Hamiltonian can be diagonalized, and reduces to

\begin{equation}
\begin{aligned}
H(\theta_n=\theta) 
&= \sum_{k} 
\begin{pmatrix}
c^\dagger_{k,B} & c^\dagger_{k,A}
\end{pmatrix} 
\mathcal{H}(k)
\begin{pmatrix}
c_{k,B} \\
c_{k,A} \\
\end{pmatrix},
\end{aligned}
\label{Eq: hamil_mom}
\end{equation} 
where, 
\begin{align*}
\mathcal{H}(k)=
\begin{pmatrix}
-2t_B \cos(k+\theta) & -t'_{AB} \\
- t'_{AB} & -2t_A \cos k
\end{pmatrix},
\end{align*}    
and $c_{k,A}=\frac{1}{\sqrt{N}}\sum_{m\in A}e^{-ikm}c_m$ and $c_{k,B}=\frac{1}{\sqrt{N}}\sum_{n\in B}e^{-ikn}c_n$ are momentum space fermionic annihilation operators 
(essentially, they are the discrete Fourier transformation of real space fermionic annihilation operators corresponding to leg A and leg B). Certainly, if $\theta$ is site-dependent, the Hamiltonian $H$ is no longer diagonal in the momentum basis. However, if $\theta$ varies slowly with position, this may still provide a good approximation. Unfortunately, this is not the case for a quasiperiodic spatial variation of $\theta$ like ours. Nevertheless, to examine whether the semiclassical approximation captures the qualitative features of the full quantum model, we express the Hamiltonian $\mathcal{H}$ in the semiclassical limit as follows:
\begin{align*}
    \mathcal{H}_{cl}(k,x)=
    \begin{pmatrix}
-2t_B\cos(k+\theta(x)) & -t'_{AB}\\
-t'_{AB} & -2t_A\cos k
\end{pmatrix}. 
\end{align*}
Here, $k$ and $x$ are no longer operators, but ordinary classical dynamical variables corresponding to momentum and position, respectively. The question we ask is whether this classical Hamiltonian $\mathcal{H}_{\mathrm{cl}}$ can exhibit signatures of localization and delocalization in the usual phase-space trajectories, in a way that can be treated analytically? If so, the hope is that this may provide some understanding of the numerical results previously obtained for the quantum lattice Hamiltonian. Note that similar semiclassical approximations of quantum lattice models have been studied previously with reasonable success in Ref.~\cite{morice2022quantum,rajbongshi2025topological}, where the hopping amplitudes also vary from site to site.

Now, $\mathcal{H}_{cl}$ is a two-band Hamiltonian. We set $t_A=t_B=t'_{AB}=1$,  and obtain by diagonalizing $\mathcal{H}_{cl}$
 two energy bands as,
\begin{equation}
 \begin{aligned}
   E_{\pm}=-\big(\cos(k+\theta)+\cos k\big)  \pm \sqrt{\big(\cos(k+\theta)-\cos k\big)^2 +1}.\\
\end{aligned}
\label{Eq: energy eq}
\end{equation}
We write the dynamical equation of motion as,

\begin{equation}
\begin{aligned}
\dot{x}=\frac{\partial E_{\pm}}{\partial k}=&\sin(k+\theta) +\sin k\\
    &\pm \frac{[\cos(k+\theta)-\cos k][-\sin(k+\theta)+\sin k]}{\sqrt{[\cos(k+\theta)-\cos k]^2 +1}},\\
 \dot{k}=-\frac{\partial E_{\pm}}{\partial x}=&-\sin(k+\theta)\frac{\partial{\theta}}{\partial{x}} \\
    &\pm \frac{[\cos(k+\theta)-\cos k]\sin(k+\theta)\frac{\partial{\theta}}{\partial{x}} }{\sqrt{[\cos(k+\theta)-\cos k]^2 +1}}.
\end{aligned} 
\label{Eq: dynamical classical eq}
\end{equation}
Given initial conditions $x = x_i$ and $k = k_i$, one expects that if $|x(t) - x_i|$ grows without bound as time evolves, it indicates delocalization in real space. Conversely, if $|x(t) - x_i|$ remains bounded for all time, it indicates localization.

\subsection{Uniform magnetic flux}
It is straightforward to verify from Eq.~\eqref{Eq: dynamical classical eq} that, in the case of a uniform magnetic flux, i.e., $\theta(x) = \theta$, one has $\dot{k} = 0$, implying that $k(t) = k_i$ remains constant, while $x(t) \sim t$. This corresponds to ballistic spreading for any value of $\theta$. Consequently, we do not expect any signature of localization.  
This absence of localization has also been observed in the quantum lattice model for constant $\theta$. In this sense, the semiclassical prediction is in agreement with the results of the quantum lattice model.

\subsection{Quasiperiodic magnetic flux}
We now consider a quasiperiodic magnetic flux of the form
\[
\theta(x)=\frac{V\pi \cos(2\pi\beta x)}{1-\lambda \cos(2\pi \beta x)},
\]
and focus on the regime $0\le\lambda<1$. 
In contrast to the case of constant magnetic flux, solving the dynamical Eq.~\eqref{Eq: dynamical classical eq} analytically for such a $\theta(x)$ is highly non-trivial. Therefore, we turn to a stability analysis of Eq.~\eqref{Eq: dynamical classical eq}.
As a first step, we determine the fixed points, namely the values of $x = x_0$ and $k = k_0$ for which $\dot{x} = \dot{k} = 0$.

\noindent One can easily find the fixed points of the dynamics,
\begin{align*}
    k_0=n\pi,~x_0=\frac{1}{2\pi\beta}\cos^{-1}\bigg[\frac{m}{V+m\lambda}\bigg].
\end{align*}
where $|\frac{m}{V+m\pi} |\le 1$.

\noindent Another set of fixed points are:
\begin{align*}
     k_0=n\pi-\frac{(-1)^m V\pi/2}{1-(-1)^m\lambda},~x_0=\frac{m\pi}{2\pi\beta}. 
\end{align*}
Lastly, another set of fixed points are:
\begin{align*}
    x_0=&\frac{m\pi}{2\pi\beta},\\
    k_0=&-\frac{\theta}{2} \pm \frac{1}{2}\cos^{-1}\big[\cos(\theta) +\frac{1}{2} \cot^2(\theta/2)\big] +n\pi.
    \end{align*}
Note that $n,~m=0,\pm1, \pm2,\cdots$.

As a next step, we linearize of the equations of motion $\dot{x}=f(x,k)$ and $\dot{k}=g(x,k)$ [here, $f(x,k)$ and $g(x,k)$ are functions corresponding to RHS of Eq.~\eqref{Eq: dynamical classical eq}] about the fixed points $(x_0,k_0)$, and yield the equations governing small fluctuations around the fixed point. Defining $\delta x = x - x_0$ and $\delta k = k - k_0$, the corresponding linearized equations of motion can be written as,\\
\begin{eqnarray}
 \begin{pmatrix} \dot{\delta x} \\ \dot{\delta k}\end{pmatrix}= \begin{pmatrix}\frac{\partial f}{\partial x}& \frac{\partial f}{\partial k} \\
\frac{\delta g}{\delta x} & \frac{\delta g}{\delta k}\end{pmatrix}_{(x_0, k_0)}\begin{pmatrix} {\delta x} \\ {\delta k}\end{pmatrix}=J \begin{pmatrix} {\delta x} \\ {\delta k}\end{pmatrix}. 
\end{eqnarray}

The stability of a fixed point is determined by the nature of the eigenvalues $\pm \omega$ of the Jacobian matrix $J$ (note since, it's a Hamiltonian system, the eigenvalues will be always in a pair of $\pm \omega$). These eigenvalues are obtained from the characteristic equation,
\begin{align*}   \omega^2=\bigg[\bigg(\frac{\partial^2 E_{\pm}}{\partial x \partial k}\bigg)^2-\frac{\partial^2 E_{\pm}}{\partial x^2}\frac{\partial^2 E_{\pm}}{\partial k^2}\bigg]_{x_0, k_0}.
\end{align*}
It is straightforward to analyze the two classes of fixed points corresponding to $\omega^2 > 0$ and $\omega^2 < 0$. When $\omega^2 < 0$, the quantity $\omega$ is purely imaginary, which implies that the eigenvalues are purely imaginary. This corresponds to stable fixed points with periodic orbits.
On the other hand, when $\omega^2 > 0$, the eigenvalues are real and of opposite signs, one positive and one negative, indicating that the fixed point is a saddle.

Figure~\ref{classical_GAA} clearly identifies the different types of fixed points for various values of $\lambda$ corresponding to our choice of $\theta(x)$. Red circles represent stable fixed points, while blue stars denote saddle points. Moreover, the red dotted lines indicate periodic orbits, and the black solid lines represent separatrices, which asymptotically connect different saddle points and separate distinct classes of trajectories. For our purposes, we will primarily focus on three saddle points, denoted by $A$, $B$, and $C$, and study their evolution as $\lambda$ varies.

For $\lambda = 0.2$, we find that the configuration of these saddle points in phase space gives rise to an unbounded region, within which trajectories can extend indefinitely in the $x$-direction, indicating delocalization. This region is denoted by the shaded area, with dashed lines highlighting the trajectories that are unbounded in the $x$-direction. With further increase in $\lambda$, the unbounded region gradually shrinks and eventually disappears. One way this can occur is if a separatrix asymptotically connects the saddle points $A$, $B$, and $C$. In that case, the shaded region corresponding to unbounded motion can no longer persist.

This is precisely what happens at $\lambda = \lambda_c$. Since the system is Hermitian and energy is conserved, all points along a given trajectory have the same energy. Therefore, if a separatrix connects $A$, $B$, and $C$, it follows that the energies corresponding to these points must be equal, i.e.,
\begin{align*}
    E_A(\lambda_c, V_c) = E_B(\lambda_c, V_c) = E_C(\lambda_c, V_c).
\end{align*}
The energy can be computed from Eq.~\eqref{Eq: energy eq}, and this condition holds independently for both bands.

 As an illustrative example, let us consider the upper energy band. Suppose the points $A$, $B$, and $C$ correspond to
\begin{align*}
A &: \left(x_0=-\frac{1}{2\pi\beta}\cos^{-1}\!\left(\frac{1}{V_c+\lambda_c}\right),\, k_0=0\right),\\
B &: \left(x_0=0,\, k_0=-\frac{V_c\pi}{2(1-\lambda_c)}\right),\\
C &: \left(x_0=-\frac{1}{2\pi\beta}\cos^{-1}\!\left(\frac{1}{V_c+\lambda_c}\right),\, k_0=-\pi\right).
\end{align*}
The critical values $V_c$ and $\lambda_c$ are then obtained by requiring that the upper-band energies at these three points become identical, i.e.,
\begin{align*}
E_{+}& \!\left(
-\frac{1}{2\pi\beta}\cos^{-1}\!\left(\frac{1}{V_c+\lambda_c}\right),\,0
\right)
=
E_{+}\!\left(
0,\,-\frac{\pi V_c}{2(1-\lambda_c)}
\right) \\
&=
E_{+}\!\left(
-\frac{1}{2\pi\beta}\cos^{-1}\!\left(\frac{1}{V_c+\lambda_c}\right),\,-\pi
\right).
\end{align*}

Above energy constraint implies the transition line in the $\lambda-V$ plane (in the regime $0\le\lambda<1$), 
\begin{align*}
    \lambda_c=1-\frac{\pi V_c}{2\cos^{-1}[\frac{1-\sqrt{5}}{2}]}.
\end{align*}
This implies that for $V_c = 1$, the critical value is $\lambda_c \simeq 0.298$, as reported in Fig.~\ref{classical_GAA}. Upon further increasing $\lambda$, the fixed points reorganize themselves. However, the unbounded shaded region observed for $\lambda < \lambda_c$ does not reappear, indicating that localization in phase space persists for $\lambda > \lambda_c$. Moreover, in the limit $\lambda_c\to 1$ ($\lambda_c\to 0$), $V_c\to 0$ ($V_c\to 1.43$).  Hence, even within our semiclassical analysis, $V_c$ decreases with increasing $\lambda_c$, qualitatively reproducing the behavior observed in the phase diagram shown in Fig.~\ref{phase_diag}. 

Note that, although our semi-classical analysis does not predict the exact transition point as in the quantum lattice, nor can it identify the intermediate phase separately, it efficiently predicts the transition in the case of a quasiperiodic magnetic flux (and absence of it for uniform flux). 

\section{Conclusion} \label{conclusion}
In this work, we aim to control single-particle transport in a low-dimensional system through the modulation of magnetic flux. We find that a strictly one-dimensional tight-binding lattice with open boundary conditions is insensitive to such flux modulation. To overcome this limitation, we introduce a two-leg ladder lattice, where the magnetic flux affects the eigenstate properties in a nontrivial manner. While a constant flux fails to induce localized states in the spectrum, a completely random magnetic flux immediately generates localization across the entire spectrum, thereby halting single-particle transport. We show that a generalized quasiperiodic modulation of the magnetic flux offers enhanced flexibility, allowing controlled tuning of single-particle transport. This allows us to access a parameter space that not only exhibits a delocalization–localization transition, but also hosts an intermediate phase.  
Moreover, this intermediate phase supports sub-ballistic transport, with the dynamical exponent spanning a broad range, $0 < \gamma < 2$. This provides significant control over single-particle transport properties in low-dimensional quantum systems, in contrast to systems that host either only a completely delocalized phase or a completely localized phase.

Moreover, systems of ultracold atoms in optical lattices provide an excellent platform for the experimental realization of models such as ours~\cite{exp_ladder_18}. In future studies, it will be interesting to investigate the many-body dynamics and the role of interactions in similar systems. In particular, it would be highly fascinating to explore how the single-particle intermediate phase is modified in the presence of different types of interactions.

\begin{acknowledgments}
R.M. acknowledges the DST-Inspire research grant from
the Department of Science and Technology, Government of India.
\end{acknowledgments}

\bibliography{apssamp}

\appendix
\section{Finite size scaling of $\langle NPR \rangle$ and $\langle IPR \rangle$}\label{finte_size_scaling}

\begin{figure}
    \centering
    \includegraphics[width=0.5\textwidth]{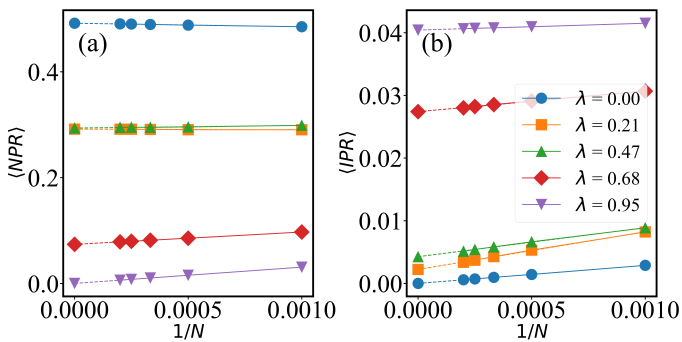}
    \caption{ Finite size extrapolation of (a) $\langle NPR \rangle$ (b) $\langle IPR \rangle$ for $V=1,$ at different $\lambda$.}
    \label{fig:ipr_npr_scal}
\end{figure}

As discussed in Sec. \ref{diagn_tools}, we classify phases of a system based on the values of $\langle NPR \rangle$ and $\langle IPR \rangle$. To obtain results in the thermodynamic limits, we perform finite-size scaling of these quantities. This is done in the following way: we first calculate these quantities for several system sizes (2$N$) with fixed values of the parameters ($V$ and $\lambda$). We then plot them as a function of $1/N$ and perform a polynomial fit to extract their values in the thermodynamic limit ($N\to \infty$). This is illustrated in Figs. \ref{fig:ipr_npr_scal}(a) and \ref{fig:ipr_npr_scal}(b) for $V=1$ and some specific values of $\lambda$.

\section{Lyapunov exponent: $V=1$ and $0\le \lambda<1$} \label{lya_expo}
\begin{figure*}[t]
    \centering
    \includegraphics[width=0.9\linewidth]{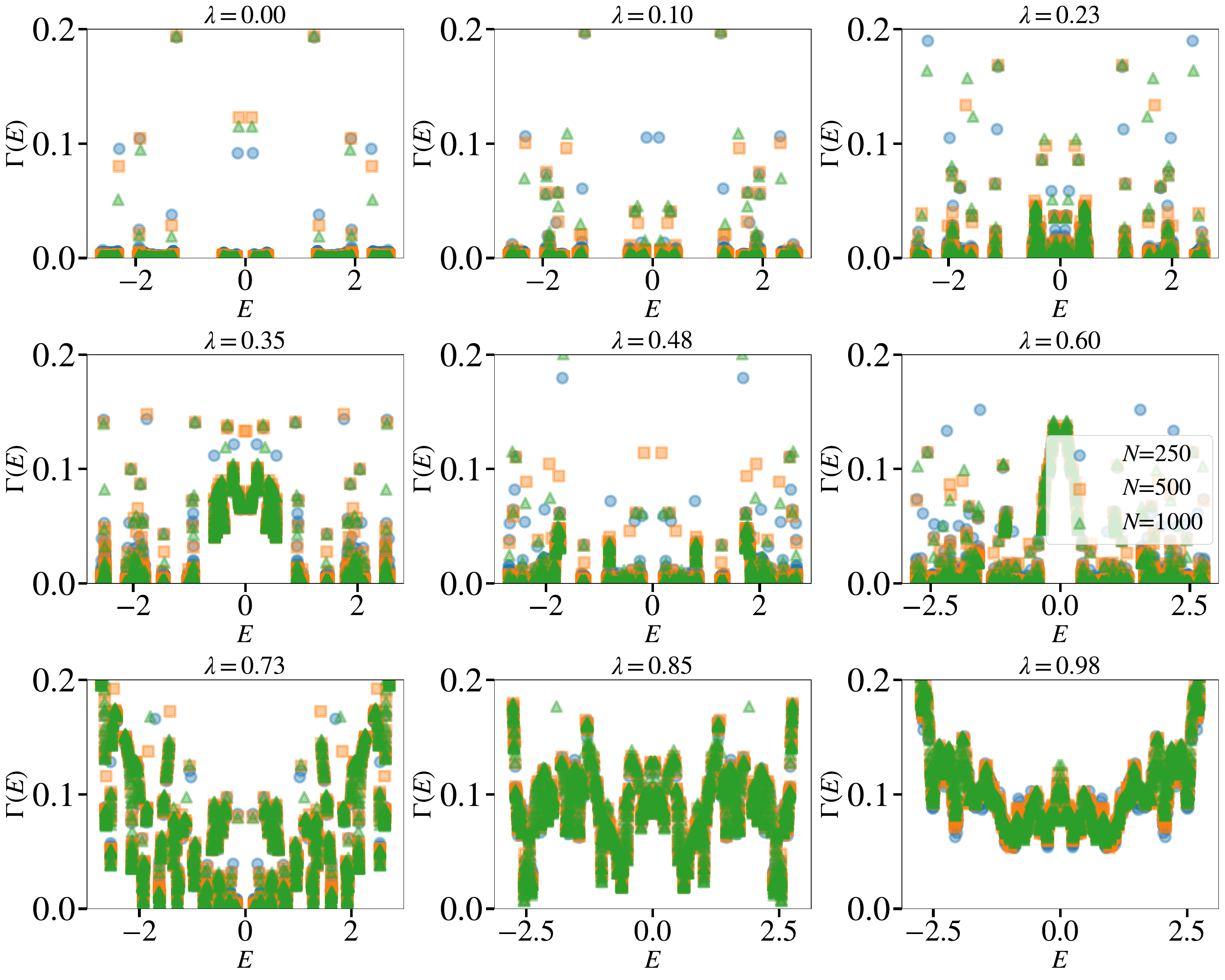}
    \caption{Quasiperiodic magnetic flux ($V=1$): $\Gamma(E)$ vs. $E$ plot for different values of $\lambda$ for $N=250, 500, 1000.$}
    \label{fig:gamma_E}
\end{figure*}
Here, we explicitly study the energy-dependent Lyapunov exponent $\Gamma(E)$ obtained using Eq.~\eqref{LE} in the main text for $V=1$ and $\lambda < 1$. Since the inverse of $\Gamma(E)$ can be interpreted as the localization length, one expects that for localized states $\Gamma(E)$ remains finite and does not scale with $N$. On the other hand, for delocalized states, $\Gamma(E) \to 0$ as $N \to \infty$.

In the main text, the NPR contour plot (see Fig.~\ref{cont_plt}) already suggests that for small $\lambda \ll 1$, almost all states are delocalized. Around $\lambda \simeq 0.35$, a significant number of mid-spectrum states appear to be localized. This is confirmed by our analysis of $\Gamma(E)$ in Fig.~\ref{fig:gamma_E}: for small $\lambda$, $\Gamma(E)$ for most states approaches zero as $N$ increases, whereas for $\lambda \simeq 0.35$, $\Gamma(E)$ corresponding to mid-energy states remains finite and does not change with $N$.

Moreover, for $\lambda = 0.85$ and $0.98$, almost no states remain delocalized, as $\Gamma(E)$ is finite for all states and does not scale with $N$. This confirms our finding in the main text that for $\lambda \gtrsim 0.8$, the phase is completely localized (see the inset of Fig. \ref{pr vs L V1}).

\section{Memory of initial state: $V=1$ and $0\le \lambda<1$}\label{memory}
\begin{figure}
    \centering
\includegraphics[width=0.48\textwidth]{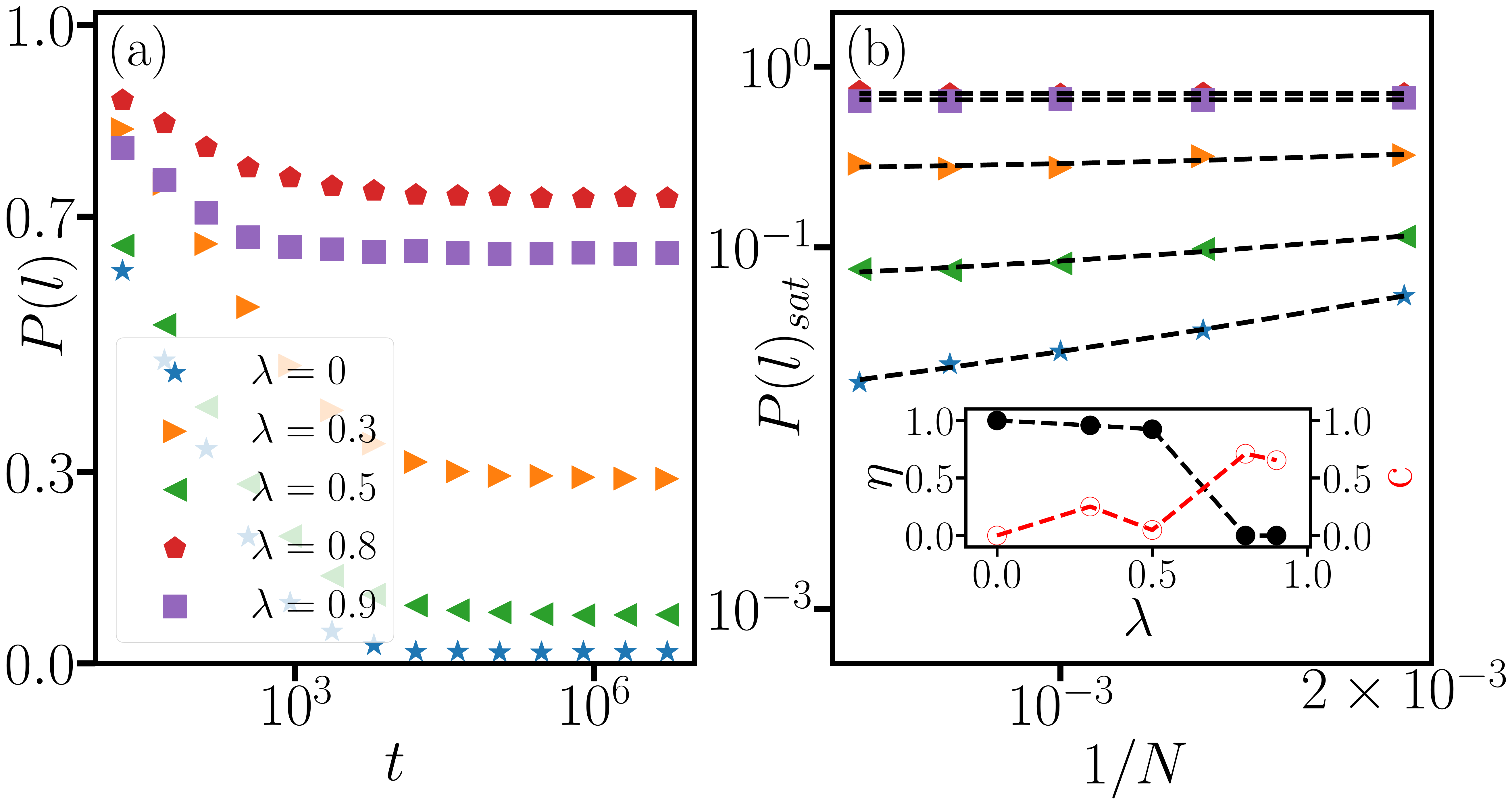}
\caption{Quasiperiodic magnetic flux ($V=1$): (a) $P(l)$ vs. $t$ for different values of $\lambda$ (for $N=1500$ and $l=10$). (b) $P(l)_{sat}$ vs. $1/N$. The black dashed line represents the fitting of the data with the functional form $P(l)_{sat} = c_0/N^{\eta} + c$. The inset shows the variation of $\eta$ and $c$ with $\lambda$.}
\label{dy_GAA_memory_1}
\end{figure}

Here, we  study the memory of an initial state by analyzing the following quantity:  
\begin{equation}\label{pl_memo}
    P(l,t)= \sum_{m=N/2-l}^{N/2+l} |\psi_A(m,t)|^2,
\end{equation}
where $l \ll N$. This quantity represents the total probability of finding the particle at time $t$ within a region of width $2l$ centered around its initial position. For the chosen initial state and irrespective of the parameter regime, we have $P(l,0)=1$. This measure provides information about how much memory of the initial state is retained during the time evolution.

In a completely localized phase, one expects $P(l,t)\approx 1$ for all times, provided that $l$ is larger than the typical localization length. In contrast, in a completely delocalized phase, the initial wave packet will spread out of the small region of width $2l$ and, after a sufficiently long time, the probability distribution will become approximately uniform, $|\psi_A(m,t)|^2\sim 1/N$ for all $m$. Consequently, $P(l,t)$ will decrease with time and take a value close to $2l/N$ in the long time limit. 

To better understand the correlation between the nature of a phase and the amount of memory retained from the initial state, we can define a saturation value the following way: 
\begin{equation}\label{pl_sat}
    P(l)_{\text{sat}} = \lim_{T \to \infty} \frac{1}{T-T_0} \int_{T_0}^{T} P(l,t)\, dt.
\end{equation}
In the localized phase, $P(l)_{\text{sat}}\approx 1$, while in the delocalized phase, $P(l)_{\text{sat}}\approx 2l/N$, which vanishes in the thermodynamic limit ($N\to \infty$). This indicates a complete loss of memory of the initial state. In general, $P(l)_{\text{sat}}$ obeys the following finite-size scaling relation:
\begin{equation}
    P(l)_{\text{sat}} =  \frac{c_0}{N^{\eta}}+c,
\end{equation}
where $c_0$ and $c$ are constants for given $l$. For completely localized (delocalized) phase, one expects $\eta = 0$ (1). The system retains the maximum memory of the initial state when $\eta=0$, whereas in the thermodynamic limit it retains almost no memory of the initial state when $\eta=1$.

Figure~\ref{dy_GAA_memory_1} left panel(a) shows the variation of $P(l)$ with $\lambda$. For $\lambda \geq 0.8$ (in the completely localized phase), the decay of $P(l)$ is slow and it saturates at a value not far from $1$. On the other hand, the presence of delocalized states leads to a much faster decay of $P(l)$.

We also study in Fig.~\ref{dy_GAA_memory_1} right panel(b),  the scaling of $P(l)_{\text{sat}}$ with $N$ for different values of $\lambda$. We fit the data using the function $P(l)_{\text{sat}}=c_0/N^{\eta} + c$, and the variation of the fitting parameters $\eta$ and $c$ with $\lambda$ is shown in the inset of Fig.~\ref{dy_GAA_memory_1}.

In the completely delocalized phase ($\lambda = 0$), we find $\eta \approx 1$ and $c \approx 0$, as expected. In the completely localized phase, $P(l)_{\text{sat}}$ remains nearly constant with $N$. Interestingly, for $\lambda = 0.3, 0.5, 0.8$, the scaling behavior is similar, but the fitted constant $c \neq 0$ in all cases. This indicates that a finite portion of the initial state's memory persists even in the thermodynamic limit.

\section{$\lambda=0$ and $\lambda\to1$ lines}\label{lambda_lines}

\begin{figure}[t]
    \centering
\includegraphics[width=0.44\textwidth]{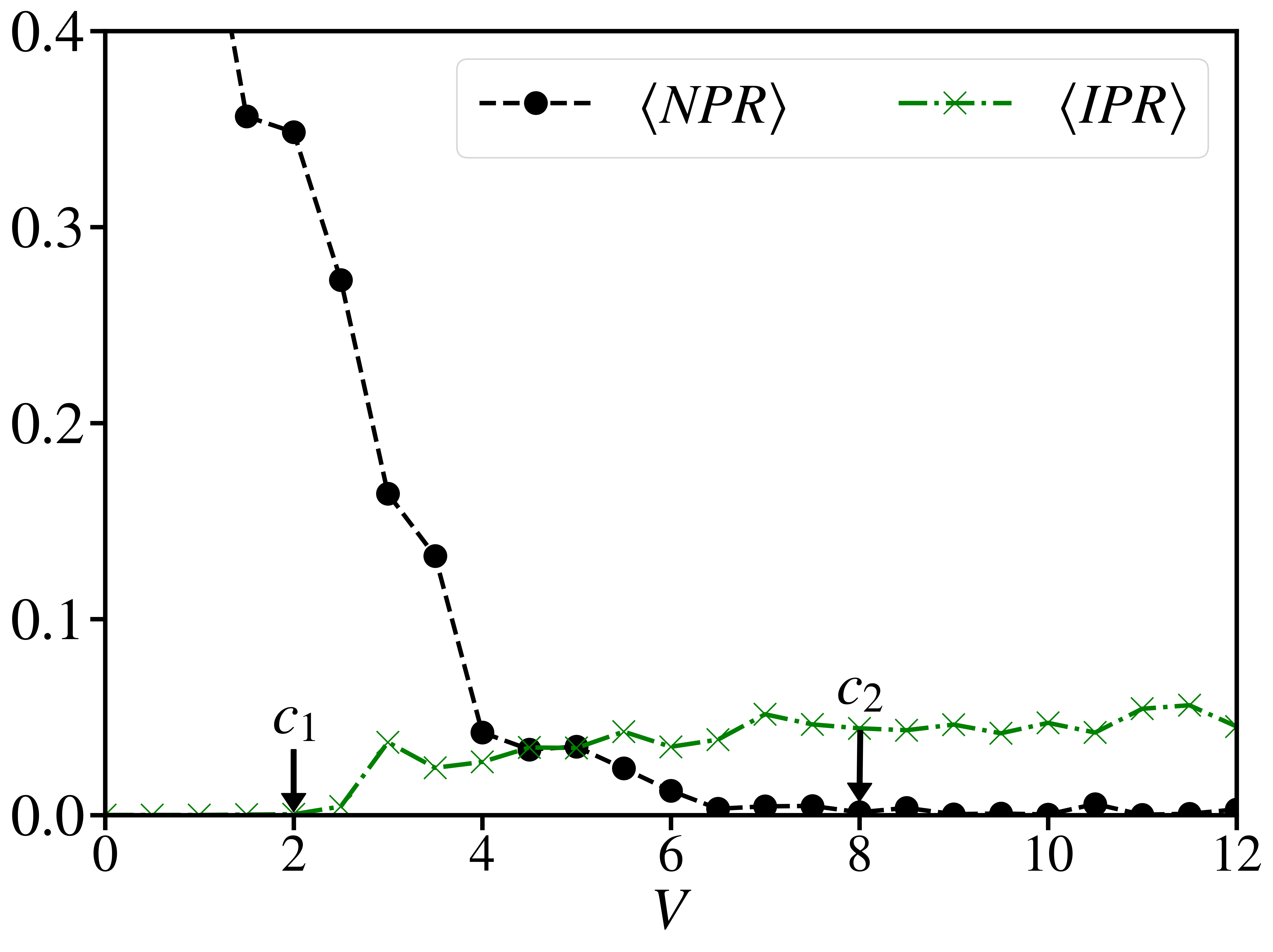}
\caption{Quasiperiodic magnetic flux ($\lambda=0$): Plot of average $IPR$ and $NPR$ as a function of $V$ in the thermodynamic limit ($N \to \infty$).}
\label{static_lam0}
\end{figure}

\begin{figure}
    \centering
\includegraphics[width=0.45\textwidth]{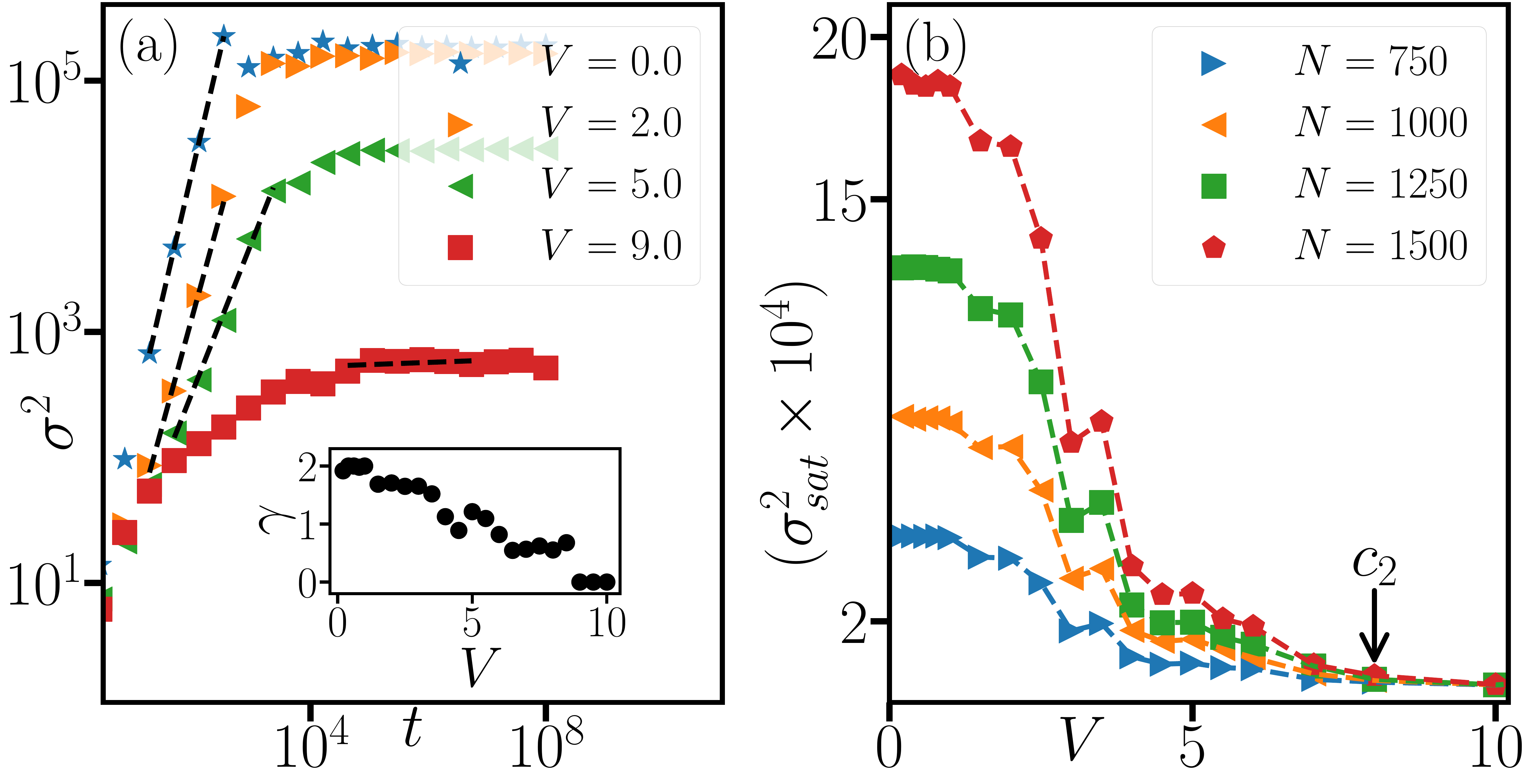}
\caption{Quasiperiodic magnetic flux ($\lambda=0$): (a) $\sigma^2$ vs. $t$ plot for different $V$ values for $N=1500$. The black dashed lines show the fitting to the functional form $\sigma^2\sim t^\gamma$. The inset shows $\gamma$ vs. $V$ plot. (b) $\sigma^2_{sat}$ vs. $V$ for different $N$ values.}
\label{dynamic_lam0}
\end{figure}

\begin{figure}
    \centering
\includegraphics[width=0.44\textwidth]{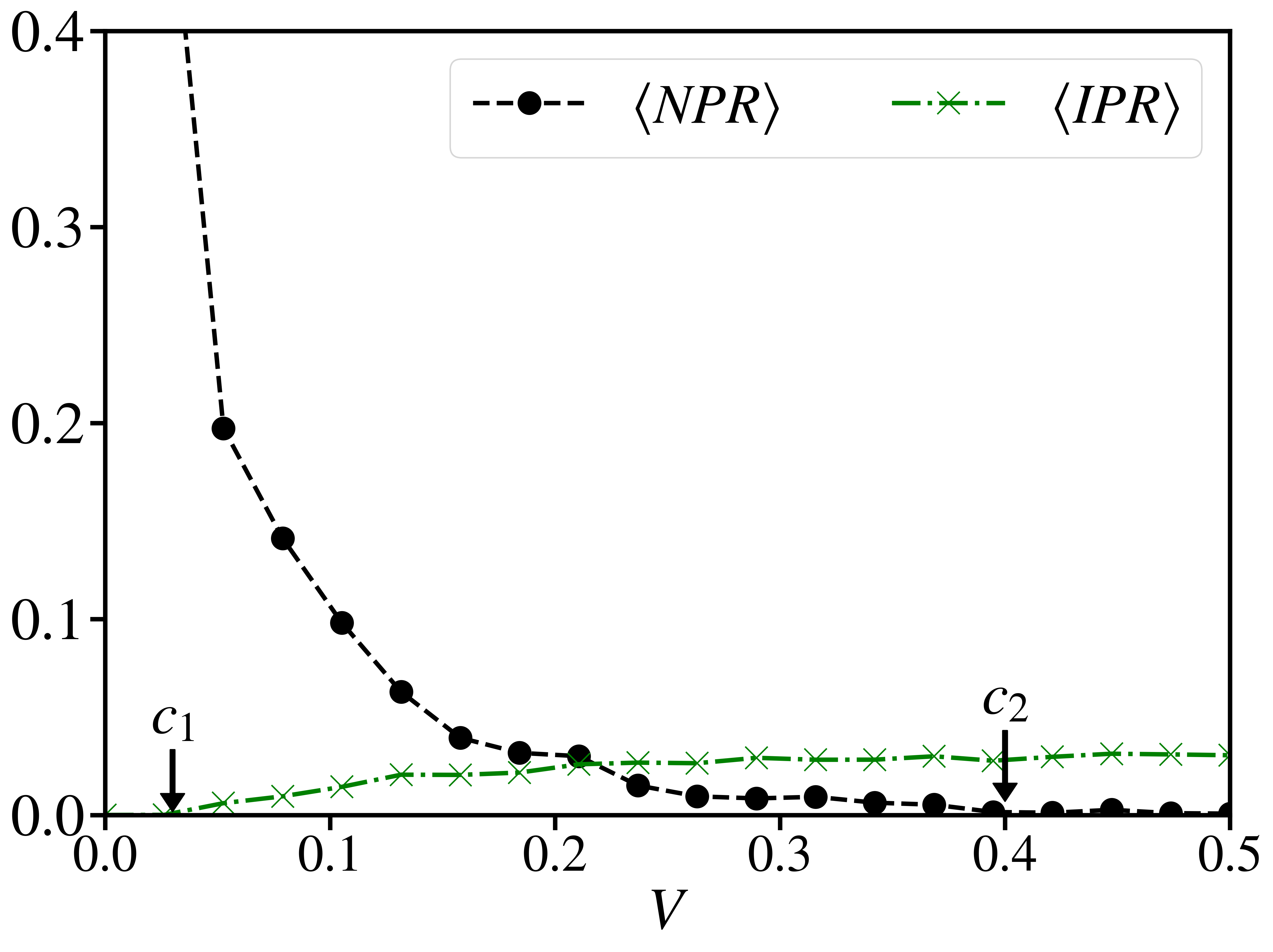}
\caption{Quasiperiodic magnetic flux ($\lambda=0.96$): Plot of average $IPR$ and $NPR$ as a function of $V$ in the thermodynamic limit ($N \to \infty$).}
\label{static_lam1}
\end{figure}

To better understand the phases along two limiting lines in our phase diagram, $\lambda=0$ and $\lambda\to 1$, we study the following static and dynamical properties.  

Along the $\lambda=0$ line, we first study $\langle NPR \rangle$ and $\langle IPR \rangle$ in the thermodynamic limit (see Fig. \ref{static_lam0}). We see that, for $0\le V\lesssim 2$, $\langle IPR \rangle =0$ and $\langle NPR \rangle > 0$. Thus, we expect the system to be in the delocalized phase in this parameter regime. Similarly, for $V\gtrsim 8$, we find that $\langle NPR\rangle = 0$ and $\langle IPR\rangle > 0$, implying that the system exhibits the localized phase in this regime. Also, we observe that both $\langle NPR\rangle > 0$ and $\langle NPR\rangle > 0$ in $2\lesssim V \lesssim 8$. We infer that the system stays in a mixed (intermediate) phase in this regime. 

We next study the single-particle dynamics along the $\lambda=0$ line. From the results presented in Fig. \ref{dynamic_lam0}, we find that $\gamma \sim 2$, $0<\gamma<2$, and $\gamma \sim 0$, respectively, for $V\lesssim 2$, $2\lesssim V \lesssim 8$ and $ V \gtrsim 8$. This result implies that the particle dynamics in those regimes is, respectively, ballistic, super- or sub-diffusive, and localized. The long time saturation value of the variance, $\sigma^2_{sat}$ (see Sec. \ref{diagn_tools}), shows no system size dependence for $V\gtrsim 8$, but shows different degrees of system size dependence for $V\lesssim 8$. This indicates that the system is in the localized phase when $V\gtrsim 8$, whereas it is in mixed or delocalized phase when $V\lesssim 8$. These results are in agreement with what we have observed in the study of static properties.

\begin{figure}[h]
    \centering
\includegraphics[width=0.45\textwidth]{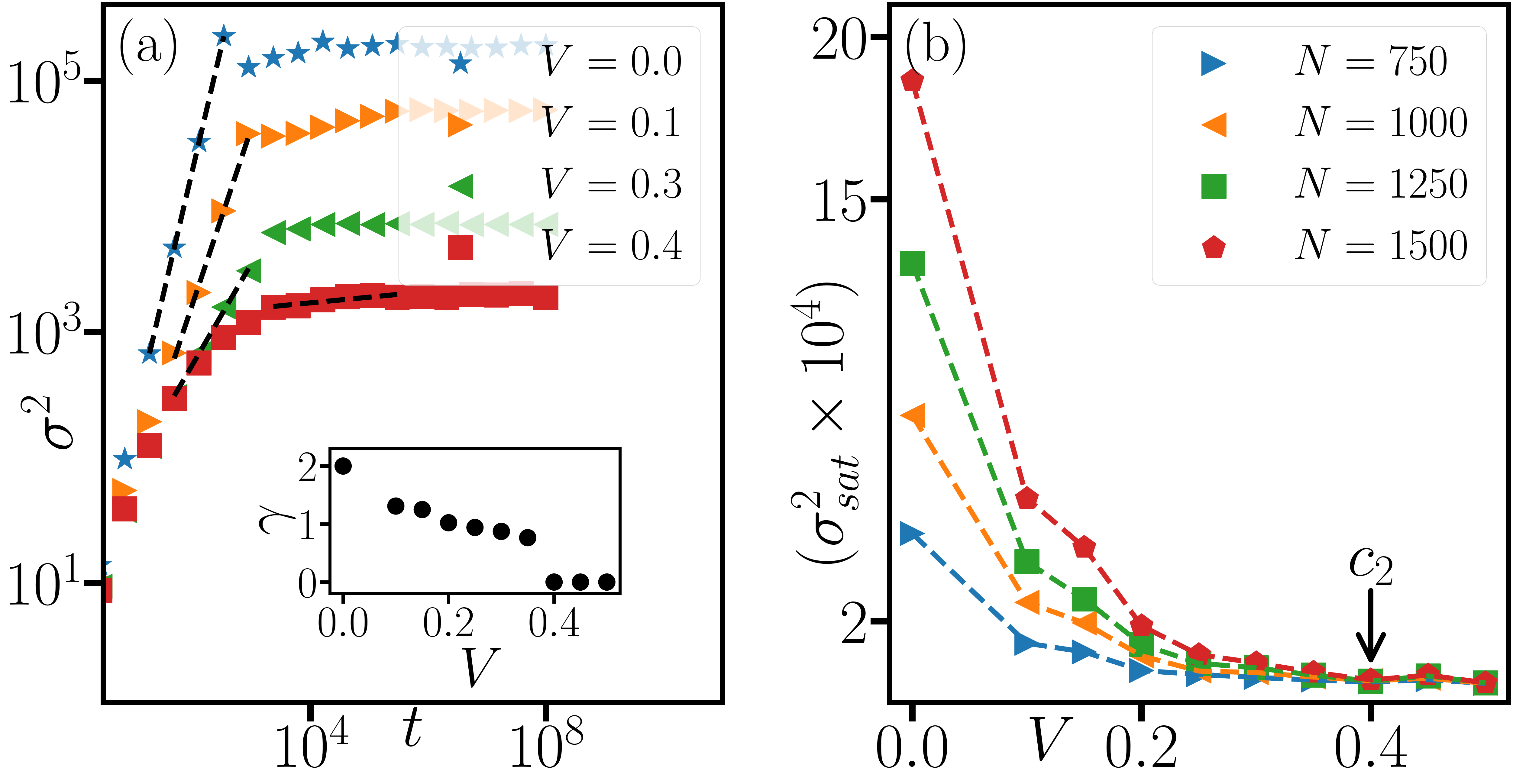}
\caption{Quasiperiodic magnetic flux ($\lambda=0.96$): (a) $\sigma^2$ vs. $t$ plot for different $V$ values for $N=1500$. The black dashed lines show the fitting to the functional form $\sigma^2\sim t^\gamma$. The inset shows $\gamma$ vs. $V$ plot. (b) $\sigma^2_{sat}$ vs. $V$ for different $N$ values.}
\label{dynamic_lam1}
\end{figure}

Similarly, a study along the $\lambda=0.96$ line is also done. The static and dynamical results can be found, respectively, in Fig. \ref{static_lam1} and Fig. \ref{dynamic_lam1}. The static results reveal that the system has the delocalized, mixed and localized phases in, respectively, $V\lesssim0.05$, $0.05\lesssim V \lesssim 0.4$ and $V\gtrsim 0.4$ regimes. The results from the dynamical study are in good agreement with these static results.

\section{Peierls phase correlation and the mechanism of reentrant localization}\label{appn_reentrant}

As observed in Fig. \ref{NPR_IPR_V_1}, our ladder system shows a reentrant localization phenomenon as function of the parameter $\lambda$. This is also evident in Figs. \ref{cont_plt} and \ref{pr vs L V1}. This phenomenon can be qualitatively understood in terms of the non-monotonic evolution of the spatial correlations in the Peierls phases with the parameter $\lambda$.

\begin{figure}[]
    \centering
\includegraphics[width=0.45\textwidth]{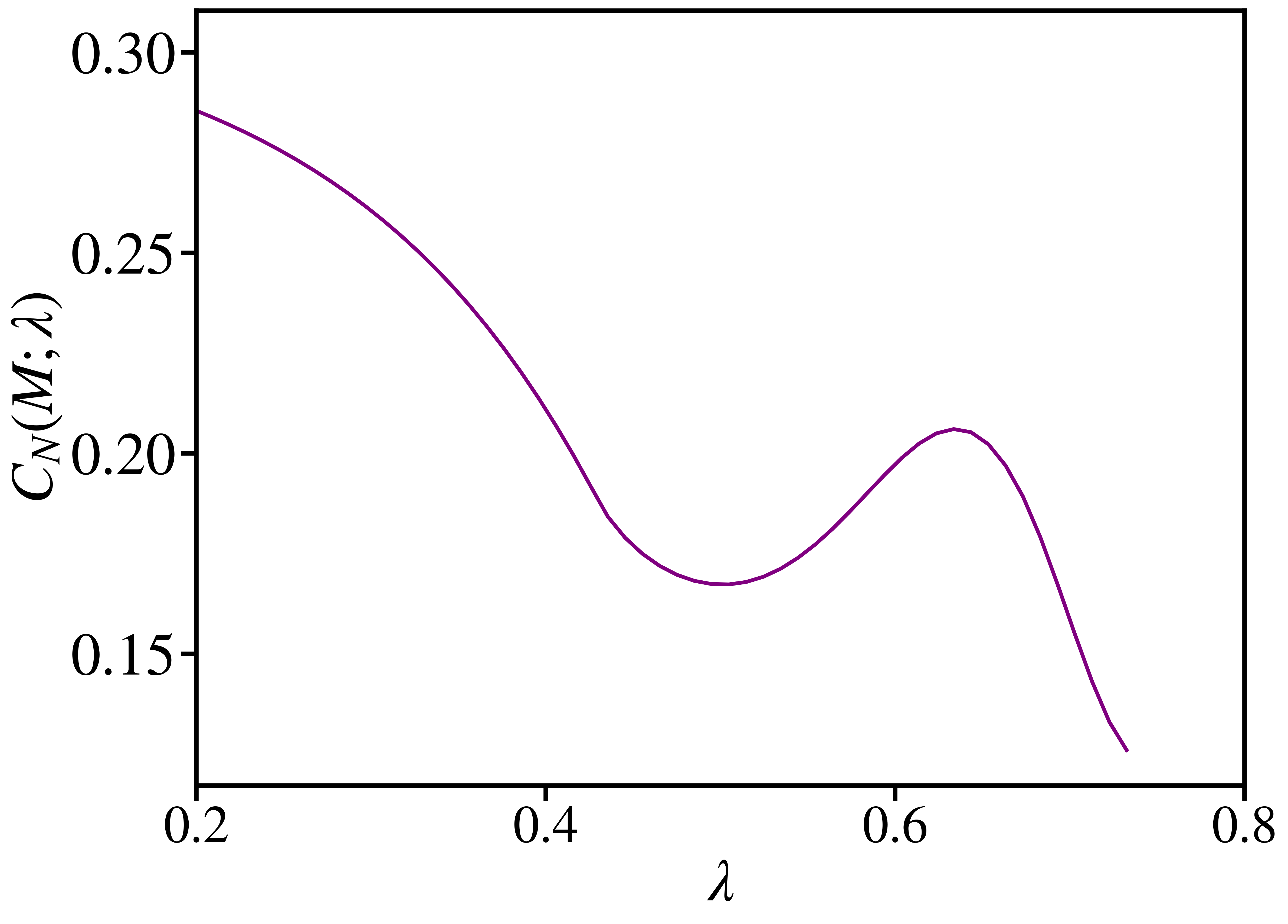}
\caption{Average spatial correlation in Peierls phase, $C_N(M;\lambda)$, as a function of parameter $\lambda$ for $V=1$. We choose $M=5000$ and $N=2000$ for the calculations.}
\label{correl}
\end{figure}

Intuitively, in the absence of spatial correlations in the disorder, one expects complete Anderson localization, as illustrated in Fig. \ref{rndm_stat} of the main text for randomly chosen Peierls phases. In contrast, strong spatial correlations, such as those arising from a periodic modulation of the Peierls phase, are expected to favor delocalization.

To quantify the spatial correlations in the Peierls phase $\theta_n$ as function of $\lambda$, we define an averaged spatial correlation function in the following way:
\begin{equation}
    C_N(M;\lambda)=\frac1N \sum_{r=1}^{N}|J_M(\lambda,r)|,
\end{equation}
where $J_M(\lambda, r)$ is the two-point correlation between the $n$-th and $(n+r)$-th phase values and is written as:
\begin{equation}
 J_M(\lambda, r) =\frac 1M \sum_{n=1}^{M} (e^{i\theta_n}-I_M(\lambda))(e^{i\theta_{n+r}}-I_M(\lambda))^*,    
\end{equation}
with 
\begin{equation}
I_M(\lambda)=\frac1M\sum_{n=1}^{M}e^{i\theta_n}.    
\end{equation}
Here Peierls phase $\theta_n$ is given by the Eq. \ref{theta_n} with $\phi=\theta=0$ and $V=1$. For thermodynamically relevant results, one should take $N,M \to \infty$. For our calculations, we choose $M=5000$ for performing statistical (or ensemble) averaging of the two-point correlation function for a fixed spatial distance $r$ and choose $N=2000$ to average over different spatial distances (the value of $N$ here reflects the typical system size we have considered). We also verified the robustness of our results by varying the values of $M$ and $N$, as well as by averaging the correlation function over at least 20 values of the phase-shift parameter $\phi$, chosen uniformly from the interval $[0,2\pi)$. In all cases, we observed no noticeable changes in the results.

The behavior of $C_N(M;\lambda)$ as a function of $\lambda$ is shown in Fig.~\ref{correl}. We find that the non-monotonic dependence of $C_N(M;\lambda)$ on $\lambda$ closely tracks the reentrant behavior observed in the localization properties of our model (see the $\langle NPR \rangle$ versus $\lambda$ plot in Fig. \ref{NPR_IPR_V_1} of the main text).

For small values of $\lambda$, the denominator in the phase modulation $\theta_n$ remains close to unity, and the Peierls phases retain a relatively smooth quasiperiodic structure across the lattice sites of leg B. Consequently, the phase correlations remain strong over long distances, resulting in a comparatively large value of $C_N(M;\lambda)$, consistent with a delocalized phase.

As $\lambda$ increases, $C_N(M;\lambda)$ decreases, reflecting a reduction in spatial correlations and a corresponding tendency toward localization. However, as $\lambda$ increases further in the intermediate regime ($\lambda \sim 0.5$), $C_N(M;\lambda)$ exhibits a pronounced increase coinciding with the reentrant delocalized phase. 
Upon further increasing $\lambda$, the overall value of the correlation function decreases again, signaling the eventual transition back to the localized phase.

The reentrant localization phenomenon observed in the main text can thus be qualitatively understood in terms of the non-monotonic evolution of the spatial correlations of the Peierls phase with the parameter $\lambda$.

\section{Alternative approach to obtain dynamical exponent $\gamma$}\label{appndx_finite_time}
\begin{figure}[h]
    \centering
\includegraphics[width=0.45\textwidth]{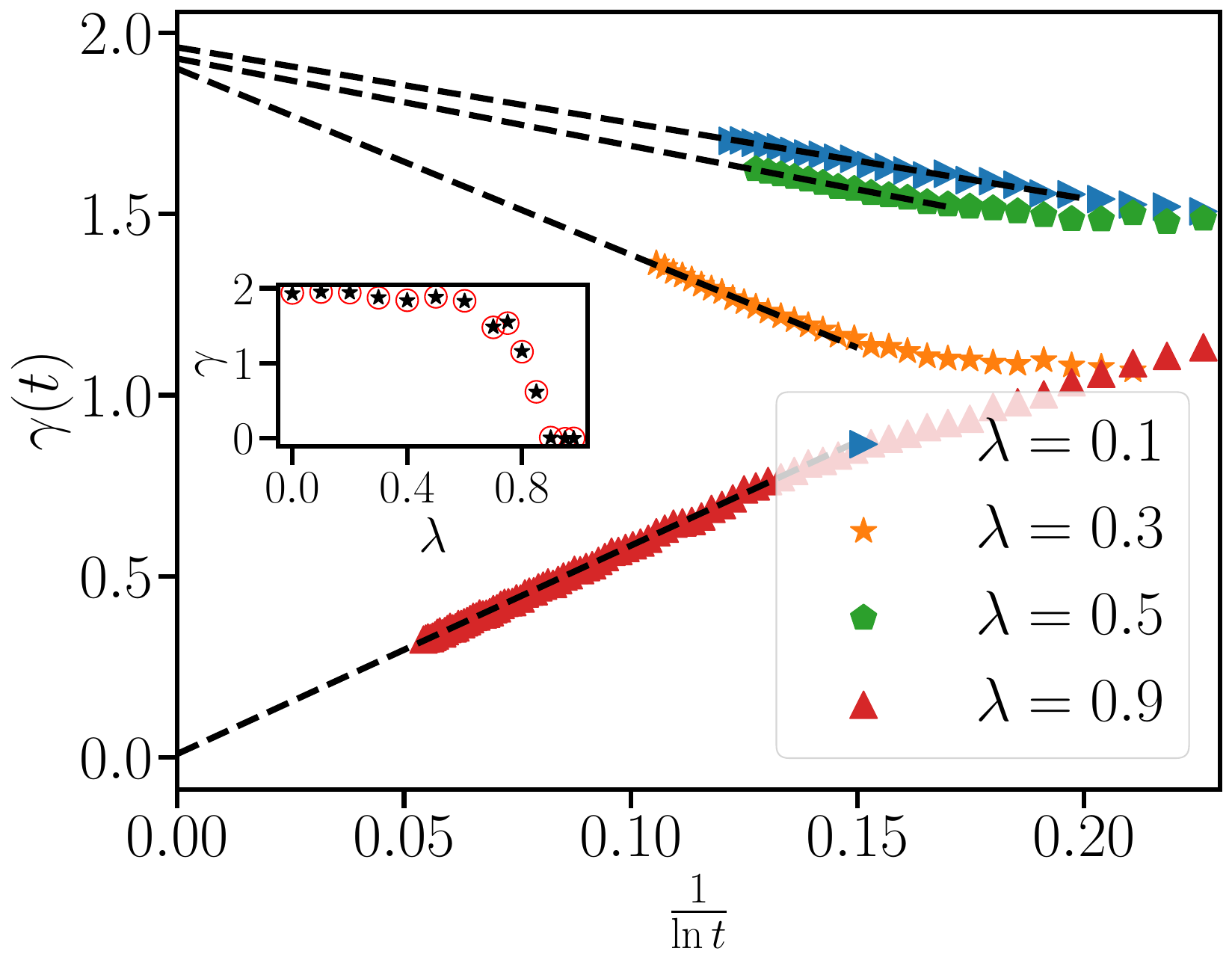}
\caption{Main panel: $\gamma(t)$ vs. $\frac{1}{\ln t}$ plot for different $\lambda$, for  $V=1$ and $N=4000$. Black dashed line represents fitting to a function $\gamma(t)=
    c_2
    +\frac{c_1}{\ln t}$. In the inset, the black star point represents the $\gamma$ value extracted from the fitting slope  $\sigma^2(t) \sim t^\gamma$, while the red circle represents the $\gamma$ value extracted from $ \gamma(t)
    =
    c_2
    +\frac{c_1}{\ln t}$. }
\label{extrapolated gamma}
\end{figure}
In this appendix,  we discuss an independent approach to extract the dynamical exponent $\gamma$ based on finite-time scaling analysis. 
In the main text, to reliably extract the dynamical exponent $\gamma$ for the delocalized or mixed regime,  we first identify a fitting time window that is both sufficiently long to capture the asymptotic scaling behavior and unaffected by finite-size saturation.  Here we take an alternative approach of finite-time scaling analysis to evaluate $\gamma$.  
By definition,
\begin{align}
    \gamma=\lim_{t\to\infty}\lim_{N\to\infty}\frac{\ln \sigma^2(t)}{\ln t}.
\end{align}
In the localized regime, $\sigma^2(t)$ remains finite even in the thermodynamic limit $N\to\infty$ and in the long-time limit $t\to\infty$. Consequently, the numerator of the above expression approaches a finite constant, while the denominator diverges, yielding $\gamma=0$.

In contrast, in the delocalized or mixed regime, $\sigma^2(t)$ diverges in the combined limits $N\to\infty$ and $t\to\infty$. Nevertheless, the ratio above remains finite and determines the dynamical exponent $\gamma$. To extract this exponent numerically, we first compute
\begin{align}
    \gamma(t)=\lim_{N\to\infty}\frac{\ln \sigma^2(t)}{\ln t}
\end{align}
for the largest accessible times in our numerics. In practice, the thermodynamic limit is approximated by using the data for $N=2500$ and $N=4000$. We identify a time window over which the data for these two system sizes coincide, indicating that finite-size effects are negligible and that the dynamics has effectively converged to the thermodynamic limit over this interval. Using these size-independent data, we compute $\gamma(t)$ and plot it as a function of $1/\ln t$ (See Fig.~\ref{extrapolated gamma}) and fit the data to
\begin{align*}
    \gamma(t)=\frac{c_1}{\ln t}+c_2.
\end{align*}
The extrapolated value
\begin{align*}
    c_2=\lim_{t\to\infty}\gamma(t)=\gamma
\end{align*}
provides an independent estimate of the asymptotic dynamical exponent $\gamma$. We find excellent agreement between this finite-time scaling analysis and the exponents obtained from direct power-law fitting of $\sigma^2(t)$ (see the inset of Fig.~\ref{extrapolated gamma}), thereby confirming the robustness of our results.

Next, we discuss why one expects the 
 leading finite-time correction to $\gamma(t)$  to scale as $1/\ln t$. 
Assuming the asymptotic form
\begin{align*}
    \sigma^2(t)=a_1t^\gamma+a_2,  
\end{align*}
and  the finite-time dynamical exponent in the $N\to \infty$ limit as
\begin{align*}
    \gamma(t)=\frac{\ln\sigma^2(t)}{\ln t},
\end{align*}
one can obtain
\begin{align*}
    \gamma(t)
    &=\frac{\ln\!\left(a_1t^\gamma+a_2\right)}{\ln t}\\
    &=\frac{\ln\!\left[a_1t^\gamma\left(1+\frac{a_2}{a_1t^\gamma}\right)\right]}{\ln t}\\
    &=\gamma+\frac{\ln a_1}{\ln t}
    +\frac{1}{\ln t}\ln\!\left(1+\frac{a_2}{a_1t^\gamma}\right).
\end{align*}
For sufficiently large $t$
one obtains, 
\begin{align*}
    \ln\!\left(1+\frac{a_2}{a_1t^\gamma}\right)
    =
    \frac{a_2}{a_1t^\gamma}
    -\frac{1}{2}\left(\frac{a_2}{a_1t^\gamma}\right)^2
    +\mathcal{O}\!\left(\frac{1}{t^{3\gamma}}\right).
\end{align*}
Substituting this into the expression for $\gamma(t)$ yields
\begin{align*}
    \gamma(t)
    =
    \gamma
    +\frac{\ln a_1}{\ln t}
    +\frac{a_2}{a_1t^\gamma\ln t}
    +\mathcal{O}\!\left(\frac{1}{t^{2\gamma}\ln t}\right).
\end{align*}
Since $t^{-\gamma}$ decays much faster than $1/\ln t$, the dominant finite-time correction is
\begin{align*}
    \gamma(t)
    =
    \gamma
    +\frac{\ln a_1}{\ln t}
    +\mathcal{O}\!\left(\frac{1}{t^\gamma\ln t}\right).
\end{align*}
Therefore, the numerical data in Fig.~\ref{extrapolated gamma} is fitted with 
\begin{align*}
    \gamma(t)=\frac{c_1}{\ln t}+c_2,
\end{align*}
where the extrapolated value
\begin{align*}
    c_2=\lim_{t\to\infty}\gamma(t)=\gamma
\end{align*}
provides the asymptotic dynamical exponent.

\end{document}